\documentclass[aps,prxquantum,twocolumn,superscriptaddress,longbibliography]{revtex4-2} 
\pdfoutput=1 
\usepackage{amsmath,amssymb,amsfonts} \usepackage{amsthm} \usepackage{physics} \usepackage{braket} \usepackage{bm} \usepackage{mathtools} 
\usepackage{graphicx} \usepackage{tikz} \usepackage{float} 
\usepackage{caption} \usepackage{subcaption} 
\usepackage{multirow} \usepackage{booktabs} 
\usepackage{algorithmicx} \usepackage{algpseudocode} 
\usepackage{xcolor} \usepackage{tcolorbox} 
\usepackage[english]{babel} 
\usepackage[colorlinks=true,linkcolor=blue,citecolor=blue,urlcolor=blue]{hyperref} 
\usepackage{enumitem} \usepackage{xcolor} \definecolor{criticalred}{HTML}{C0392B} \definecolor{highorange}{HTML}{E67E22} \definecolor{mediumpurple}{HTML}{8E44AD} \definecolor{lowgray}{HTML}{7F8C8D}     
       \renewcommand{\Im}{\operatorname{Im}} \newcommand{\Spseudo}{S_{\mathrm{pseudo}}}   \newcommand{\BZ}{\mathrm{BZ}} \newcommand{\cdag}{c^{\dagger}}  \newcommand{\tauA}{\tau^{1|2}_A} \newcommand{\tauFull}{\tau^{1|2}} \newcommand{\mK}{\mathcal{K}} \newcommand{\NA}{N_A} \newcommand{\wv}[1]{\,{}^{2}\!\langle #1 \rangle_1}  
\begin{document} 
\title{Pseudo entropy and topological phases of matter} 
\author{Pramod Kamal Kharel} 
\thanks{These authors contributed equally to this work.} 
\affiliation{Department of Physics and Astronomy, Ohio University, Athens, OH 45701, USA} 
\affiliation{Holographic Himalaya, Lambagar, 44600 Tarakeshwar, Nepal} 
\author{Manghang Limbu} \thanks{These authors contributed equally to this work.} 
\affiliation{St. Xavier’s College, Tribhuvan University, Maitighar, 44600 Kathmandu, Nepal} \affiliation{Holographic Himalaya, Lambagar, 44600 Tarakeshwar, Nepal} 
\author{Nabaraj Khatri} \thanks{These authors contributed equally to this work.} 
\affiliation{Tri-Chandra Multiple Campus, Tribhuvan University, Ghantaghar, 44600 Kathmandu, Nepal} \affiliation{Holographic Himalaya, Lambagar, 44600 Tarakeshwar, Nepal} 
\author{Ashish Khanal} \thanks{These authors contributed equally to this work.} 
\affiliation{Department of Physics and Astronomy, Ohio University, Athens, OH 45701, USA} 
\affiliation{Holographic Himalaya, Lambagar, 44600 Tarakeshwar, Nepal}
\author{Kiran Adhikari} 
\email{kiran.adhikari@tum.de} 
\affiliation{Emmy Noether Group for Theoretical Quantum Systems Design, Technical University of Munich, Arcisstraße 21, 80333 München}

\begin{abstract}
Entanglement entropy has proven to be a powerful probe of phenomena such as quantum chaos and phase transitions. Pseudo entropy is a recently proposed time-like generalization of an entanglement measure, motivated by de Sitter holography. In this work, we find that pseudo entropy can also serve as a novel probe for distinguishing topological phases of matter. For this, we consider the Su--Schrieffer--Heeger model as a representative example and investigate the averaged excess entropy $\Delta S_{12}$, defined as the difference between pseudo entropy and the average entanglement entropy, across the topological-to-trivial and trivial-to-topological phase transitions. When the two states are in the same phase, we find that $ \Delta S_{12}$ is non-positive under periodic boundary conditions, while for open boundary conditions, it is non-positive only when the system is sufficiently large. Moreover, we analyze ground-state quench protocols for topology-crossing quenches and find that the imaginary pseudo entropy tracks the critical times predicted by the Fisher zeros. 
\end{abstract}

\maketitle
\section{Introduction}\label{intro} 

In 1935, Schr\"odinger recognized entanglement as the defining feature that distinguishes quantum mechanics from classical physics~\cite{Schrödinger_1935}. Since then, entanglement has become a key quantum resource of modern quantum physics, playing crucial roles in both foundational studies and quantum technologies \cite{Eisert:2008ur, Kitaev:2005dm, Adhikari:2026srf, Adhikari_2021, 10461354}. One of the areas where entanglement has found profound applications is condensed matter and many-body physics, where entanglement entropy has emerged as an important tool for characterizing quantum phases and quantum phase transitions \cite{Vidal:2002rm, Calabrese:2004eu,  Levin:2006zz, Latorre:2003kg, Adhikari:2022whf}.

Recently, a novel quantity known as pseudo entropy has been introduced as a time-like generalization of entanglement entropy~\cite{Nakata2021}. Pseudo entropy is defined as the von Neumann entropy of a transition matrix constructed from two distinct quantum states. Since the transition matrix is generally non-Hermitian, the pseudo entropy  can take complex values. In the context of holography, its development was motivated by the need to extend the geometric interpretation of entanglement entropy to Euclidean time-dependent spacetimes, where a well-defined Euclidean minimal surface is generally absent \cite{Narayan:2022afv, Nanda:2025tid, Narayan:2026wzp, Doi:2022iyj, Doi:2023zaf}. The same set of ideas has since been extended to inflationary and other cosmological geometries \cite{Goswami:2024vfl, Limbu:2026bol, Adhikari_2022, Kamal_Kharel_2026, Adhikari:2021ked}.

Although originating in holography, pseudo entropy has recently been extended to condensed matter and many-body quantum systems \cite{Mukherjee:2022jac, He:2023eap, Mollabashi:2020yie, Mollabashi:2021xsd}. It has been used to diagnose scrambling, and spectral chaos \cite{Das:2026ifj, Das:2025fcd,Adhikari:2025zoa}, and topological entropy \cite{Nishioka:2021cxe, Caputa:2024qkk}. Furthermore,  \cite{Mollabashi:2020yie, Mollabashi:2021xsd} studied pseudo entropy in Lifshitz free scalar field theories and the Ising and XY spin models and suggested that pseudo entropy can serve as a novel quantum order parameter capable of distinguishing different quantum phases and identifying phase transitions. In this work, we analyze whether pseudo entropy serves as a diagnostic for topological phases. For this, we consider the Su--Schrieffer--Heeger (SSH) model, one of the simplest models with topological phases. The SSH model is a paradigmatic one-dimensional lattice model with alternating hopping amplitudes whose $\mathbb{Z}_2$ topological invariant is protected by chiral symmetry. We find that the pseudo entropy  tracks the topological phases under periodic boundary conditions. However, for open boundary conditions, the situation is more subtle, and finite-size effects must be considered. Extending to non-equilibrium systems, we observe that the imaginary part of pseudo entropy  provides a diagnostic of dynamical phase transitions.

This work is organized as follows. In section \ref{sec:pseudo}, we review concepts of pseudo entropy and how to compute for free-fermionic systems using the correlation matrix method. In section \ref{sec:pseudo_SSH}, we compute the pseudo entropy between ground states of two SSH Hamiltonians and examine how it behaves across the topological phase transition. The analysis suggests that pseudo entropy exhibits area-law saturation in the gapped phases and logarithmic scaling at the critical point. In section \ref{sec:pseudo_conjecture_test}, we test the non-positivity conjecture for both open boundary conditions and periodic boundary conditions. In section \ref{sec:imS-DPT}, we present evidence that the imaginary part of pseudo entropy calculated after quenching can identify a dynamical phase transition. Finally, in \ref{sec:outlook},  we conclude and mention potential future directions.


\section{Pseudo Entropy}
 \label{sec:pseudo}
Pseudo entropy can be thought of as a generalization of entanglement entropy to a transition between the two pure quantum states $|\psi\rangle$ and $|\varphi\rangle$ satisfying $\langle \varphi | \psi \rangle \neq 0$ \cite{ Nakata2021,Mollabashi:2021xsd}. First, we define the transition matrix, $\mathcal{T}^{\psi|\varphi}$, as follows:
\begin{equation}
  \mathcal{T}^{\psi|\varphi} = \frac{|\psi\rangle\langle\varphi|}{\langle\varphi|\psi\rangle},
  \label{eq:transition}
\end{equation}
which is normalized such that its trace is one. Similarly, for any $n \in \mathbb{N}^+$, $(\mathcal{T}^{\psi|\varphi})^n = \mathcal{T}^{\psi|\varphi}$, giving $\mathrm{Tr}\left[\left(\mathcal{T}^{\psi|\varphi}\right)^n\right] = 1.$ Furthermore, under the exchange of $|\psi\rangle$ and $|\varphi\rangle$, $  \mathcal{T}^{\varphi|\psi} = \left(\mathcal{T}^{\psi|\varphi}\right)^\dagger.$\\
Just like in the case of computing entanglement entropy, we divide the total Hilbert space $\mathcal{H}_{\text{tot}}$ into two parts $A$ and $B$, $  \mathcal{H}_{\text{tot}} = \mathcal{H}_A \otimes \mathcal{H}_B.$ Accordingly, the reduced transition matrix of system $A$ is defined by tracing out the system $B$, 
\begin{equation}
  \mathcal{T}^{\psi|\varphi}_A \equiv \mathrm{Tr}_B\left[\mathcal{T}^{\psi|\varphi}\right] = \mathrm{Tr}_B\left[\frac{|\psi\rangle\langle\varphi|}{\langle\varphi|\psi\rangle}\right].
  \label{eq:reduced}
\end{equation}
and pseudo entropy is defined as:
\begin{equation}
\label{eq:Pseudo_entropy}
    S\!\left(\mathcal{T}^{\psi|\varphi}_A\right) = -\mathrm{Tr}\left(\mathcal{T}^{\psi|\varphi}_A \log \mathcal{T}^{\psi|\varphi}_A\right).
\end{equation}
For the case when the initial and final states are the same, i.e. $\ket{\psi} = \ket{\varphi}$, the pseudo entropy reduces to the ordinary entanglement entropy. While equation \eqref{eq:Pseudo_entropy} does look like Von Neumann entropy, equation \eqref{eq:Pseudo_entropy} can in general be complex as $\mathcal{T}^{\psi|\varphi}_A$ is in general non-Hermitian with complex eigenvalues. Only in the special choices of the initial and final states is pseudo entropy real. This can happen, for instance, when the states arise from a real-valued Euclidean action~\cite{Mollabashi:2021xsd}, or when the transition matrix itself is real so that its eigenvalues are real ~\cite{Guo:2023tjv}.  In contrast, Von Neumann entropy is always real. Operationally, pseudo entropy can be thought of as a measure of quantum entanglement for the intermediate states between the initial and final states \cite{Nakata2021}.

For calculation purposes, it is convenient to define the $n$-th R\'enyi entropy of the transition matrix $\mathcal{T}^{\psi|\varphi}_A$ just like we define the $n$-th R\'enyi entropy of a quantum state $\rho$:
\begin{equation}
  S^{(n)}\!\left(\mathcal{T}^{\psi|\varphi}_A\right) \equiv \frac{1}{1-n}\log\mathrm{Tr}\!\left[\left(\mathcal{T}^{\psi|\varphi}_A\right)^n\right],
  \label{eq:renyi}
\end{equation}
where $(n \in \mathbb{N}^+,\, n \geq 2)$, and we can simply choose the branch of the log function: $-\pi < \mathrm{Im}[\log(z)] \leq \pi$. We call this quantity $S^{(n)}(\mathcal{T}^{\psi|\varphi}_A)$ the pseudo $n$-th R\'enyi entropy, and taking the $n \to 1$ limit, one obtains the pseudo entropy:
\begin{equation}
  S\!\left(\mathcal{T}^{\psi|\varphi}_A\right) \equiv \lim_{n\to 1} S^{(n)}\!\left(\mathcal{T}^{\psi|\varphi}_A\right).
  \label{eq:von}
\end{equation}

It is also possible to define the following real-valued quantity, which can be particularly relevant for phase transition studies. 
\begin{align}
\Delta S^{(n)} \!\left(\mathcal{T}^{\psi|\varphi}_A\right) = \frac{1}{2} \Big[ &S^{(n)} \!\left(\mathcal{T}^{\psi|\varphi}_A\right) + S^{(n)} \!\left(\mathcal{T}^{\varphi|\psi}_A\right) \nonumber \\
&-   S^{(n)}\!\left(\mathcal{T}^{\psi|\psi}_A\right) -   S^{(n)}\!\left(\mathcal{T}^{\varphi|\varphi}_A\right) \Big]
\end{align}
where, we note that $S^{(n)} \left(\mathcal{T}^{\psi|\varphi}_A\right) = S^{(n)} \left(\mathcal{T}^{\varphi|\psi}_A\right)^*$, and the latter two terms are the standard entanglement entropy for the state $\ket{\psi}$ and $\ket{\varphi}$ respectively. Taking the $n \to 1$ limit, we obtain $   \Delta S \left(\mathcal{T}^{\psi|\varphi}_A\right) =  \lim_{n\to 1} \Delta S^{(n)} \left(\mathcal{T}^{\psi|\varphi}_A\right).$
$ \Delta S \left(\mathcal{T}^{\psi|\varphi}_A\right)$ is thus nothing but the difference between the real part of the pseudo entropy and the averaged entanglement entropy:
\begin{align}
\Delta S \!\left(\mathcal{T}^{\psi|\varphi}_A\right) = &\,\text{Re} \!\left(S\!\left(\mathcal{T}^{\psi|\varphi}_A\right) \right) - \frac{1}{2} \left( S(\rho_\psi)_A +  S(\rho_\varphi)_A \right)
\end{align}
where $ S(\rho_\psi)_A$ and $S(\rho_\varphi)_A$ is the standard entanglement entropy for the state $\ket{\psi}$ and $\ket{\varphi}$ of subsystem $A$.\\ 
In \cite{Mollabashi:2021xsd, Mollabashi:2020yie}, pseudo entropy was numerically evaluated for the Lifshitz free scalar field and for Ising and XY spin models. The result showed that the difference $\Delta S \left(\mathcal{T}^{\psi|\varphi}_A\right)$ satisfies the following inequality:
\begin{equation}
    \Delta S \left(\mathcal{T}^{\psi|\varphi}_A\right) \leq 0
\end{equation}
when the state $\ket{\psi}$ and $\ket{\varphi}$ in the same phase, while if there is a quantum phase transition from  $\ket{\psi}$ to $\ket{\varphi}$, the inequality is typically violated.
\begin{figure}[tbp]
    \centering
    \includegraphics[width=\linewidth]{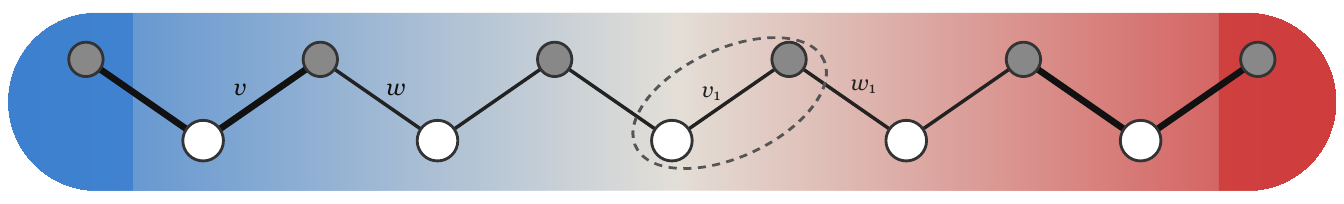}
    \caption{Lattice structure of SSH model. Filled circles are sites in subregion $A$, and empty circles are sites in $B$. They are grouped in unit cells. We have highlighted the third cell in the diagram. $v$ is the intracell hopping parameter and $ w$ is the intercell hopping parameter.}
    \label{fig:SSH_diagram}
\end{figure}

\subsection{Pseudo entropy  of Fermionic systems from Correlation Matrices}
\label{sec:Fermionic_correlator}

For a generic many-body system, computing the entanglement entropy requires diagonalizing the reduced density matrix $\rho_A$, whose dimension grows exponentially with the subsystem size. For free fermions, however, the ground state is Gaussian, and Wick's theorem implies that the two-point correlation matrix determines all reduced-state properties \cite{10.21468/SciPostPhysLectNotes.54}. For a subsystem $A$, the correlation matrix is $C_{ij}=\langle c_i^\dagger c_j\rangle, \quad i,j\in A .$
If $\{\zeta_k\}$ are the eigenvalues of $C_A$, the entanglement entropy is given by the Peschel formula \cite{IngoPeschel_2003,Peschel_2009}
\begin{equation}
S_A
=
-\sum_{k=1}^{N_A}
\left[
\zeta_k\ln\zeta_k
+
(1-\zeta_k)\ln(1-\zeta_k)
\right].
\label{eq:EE}
\end{equation}

We now briefly review the corresponding construction for pseudo entropy in free-fermion systems \cite{Mollabashi:2021xsd}; technical details are given in Appendix~\ref{apx:fermionic_pseudo}. Consider two Gaussian states $|\psi_1\rangle$ and $|\psi_2\rangle$, taken to be ground states of quadratic Hamiltonians
\begin{equation}
H^{(s)}
=
\sum_{i,j}h^{(s)}_{ij}c_i^\dagger c_j,
\qquad s=1,2,
\label{eq:general_H}
\end{equation}
where $c_i^\dagger$ creates a fermion in the $i$th single-particle orbital. After diagonalization, $H^{(s)}
=
\sum_k \epsilon_k^{(s)}
(\eta_k^{(s)})^\dagger \eta_k^{(s)},$
the ground state takes the Slater-determinant form $|\psi_s\rangle
=
\prod_{k=1}^{M_s}
(\eta_k^{(s)})^\dagger |0\rangle .$

Because both states are Gaussian, the transition matrix $\tau^{1|2}$ and its reduced transition matrix $\tau_A^{1|2}=\mathrm{Tr}_B\,\tau^{1|2}$ are also Gaussian. Consequently, the spectrum of $\tau_A^{1|2}$ can be extracted entirely from two-point weak-value correlators,
\begin{equation}
\wv{O}
=
\frac{\langle\psi_2|O|\psi_1\rangle}
{\langle\psi_2|\psi_1\rangle}.
\label{eq:wv}
\end{equation}
For the particle-number-conserving systems studied here, the anomalous correlators vanish and the pseudo-correlation matrix restricted to subsystem $A$ is simply $C^{1|2}_{ij}=\wv{c_i^\dagger c_j}, \qquad i,j\in A,$
whose eigenvalues $\{\nu_k\}$ enter equation~\eqref{eq:PE_final_main}
directly. For Slater determinants, this evaluates to $C^{1|2}\big|_A = \Bigl[U_1\bigl(U_2^\dagger U_1\bigr)^{-1}U_2^\dagger\Bigr]_A,$
where $U_s$ is the $N\times M_s$ matrix of occupied single-particle orbitals of $|\psi_s\rangle$, and $[\cdots]_A$ denotes restriction to subsystem $A$.
Following Ref.~\cite{Mollabashi:2021xsd}, one constructs a pseudo-correlation matrix from these weak values. If its eigenvalues are denoted by $\{\nu_k\}$, the pseudo entropy is obtained through the fermionic correlation-matrix formula
\begin{equation}
S(\tau_A^{1|2})
=
-\sum_{k=1}^{N_A}
\left[
\nu_k\log\nu_k
+
(1-\nu_k)\log(1-\nu_k)
\right],
\label{eq:PE_final_main}
\end{equation}
where the $\nu_k$ are generally complex because $\tau_A^{1|2}$ is non-Hermitian. The corresponding pseudo-R\'enyi entropies are computed from the same spectrum. Throughout this work, the pseudo entropy is evaluated numerically using this correlation-matrix formalism.

\begin{figure}[tbp]
    \centering
    \includegraphics[width=1\linewidth]{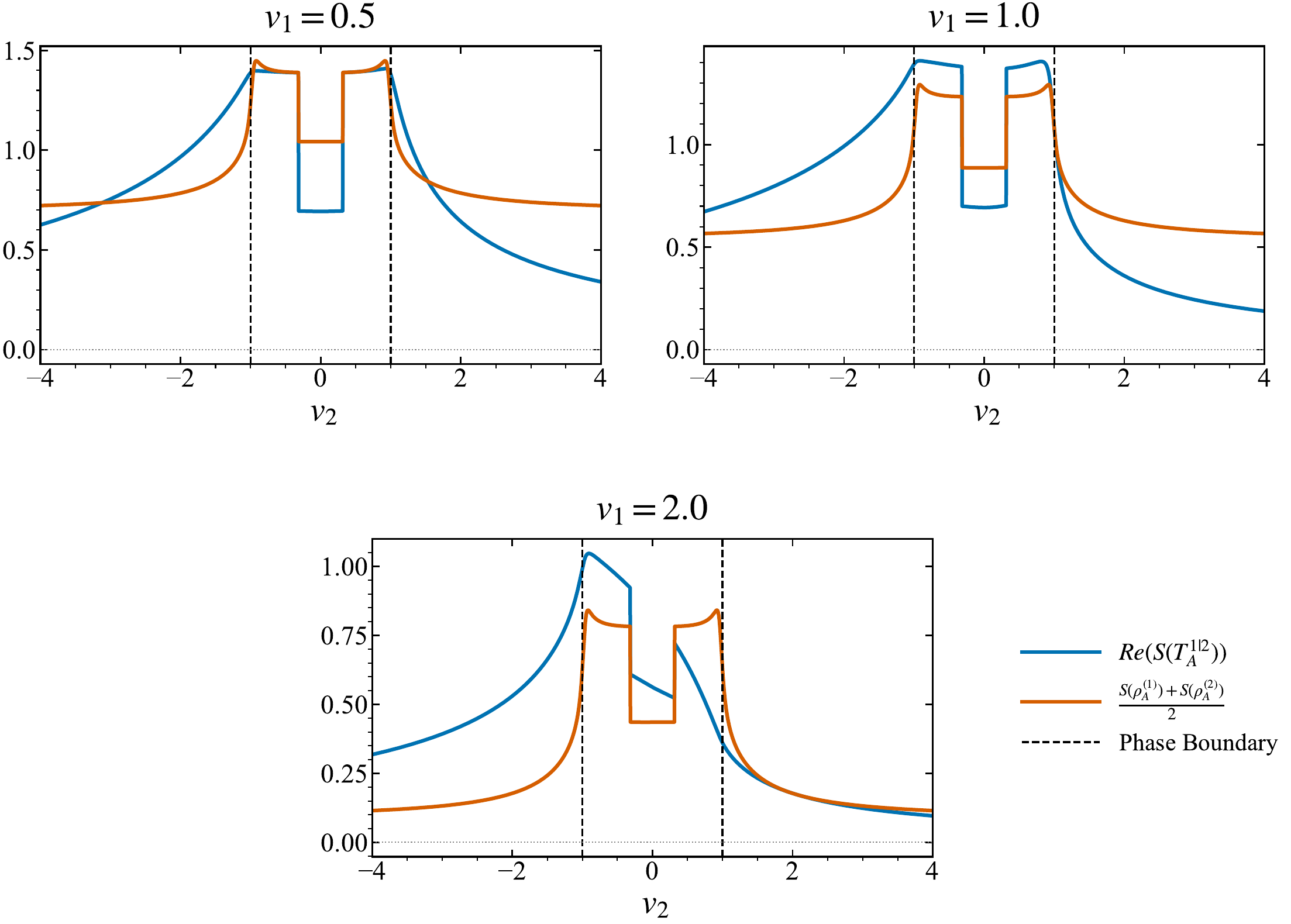}
    \caption{pseudo entropy and average entanglement entropy as a function of the hopping parameter $v_2$ for $N = 60$ unit cells, $w = 1$. The top left panel corresponds to a reference state in the topological phase ($v_1 = 0.5$), the top right panel in the critical phase boundary ($v_1 = 1.0$), and the bottom panel in the trivial phase ($v_1 = 2.0$). The vertical dashed lines at $v_2 = \pm 1$ indicate the bulk gap-closing transitions that delineate the topological regime ($\lvert v_2 \rvert < 1$) from the trivial regime ($\lvert v_2 \rvert > 1$).} 
    \label{fig:Pseudo entropy for v=0.5}
\end{figure}


\begin{figure*}[tbp]
    \centering

    \begin{subfigure}{\columnwidth}
        \centering
        \includegraphics[width=\linewidth]
        {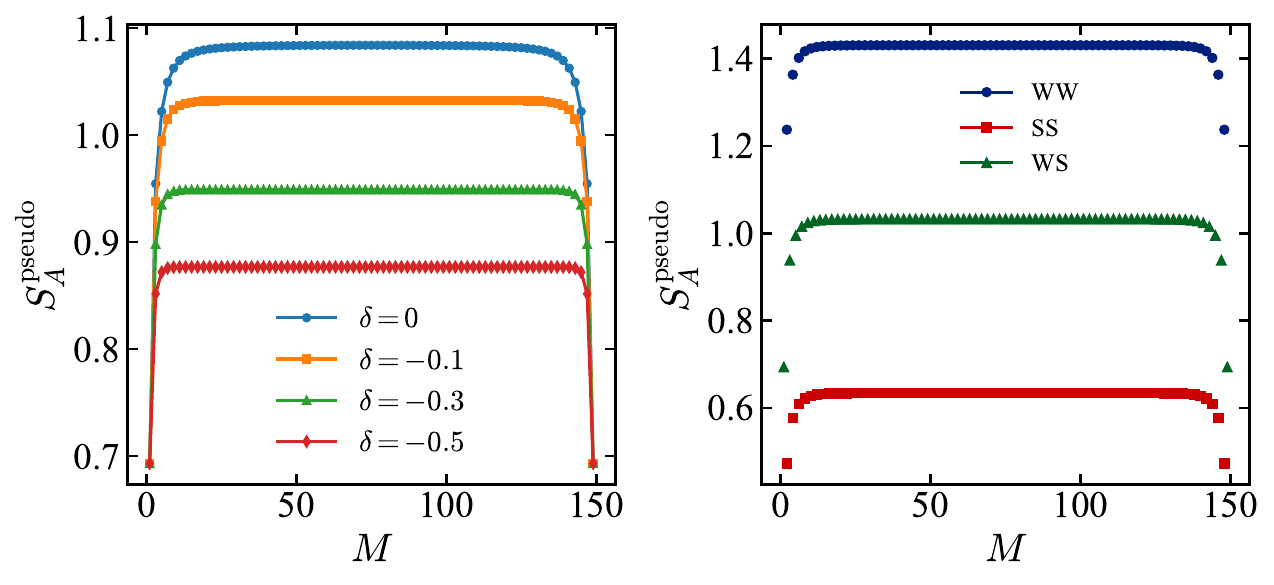}
        \caption{Topological phase, $v_{\mathrm{ref}}=0.5<w=1$.}
        \label{fig: 0.5 PES}
    \end{subfigure}
    \hfill
    \begin{subfigure}{\columnwidth}
        \centering
        \includegraphics[width=\linewidth]
        {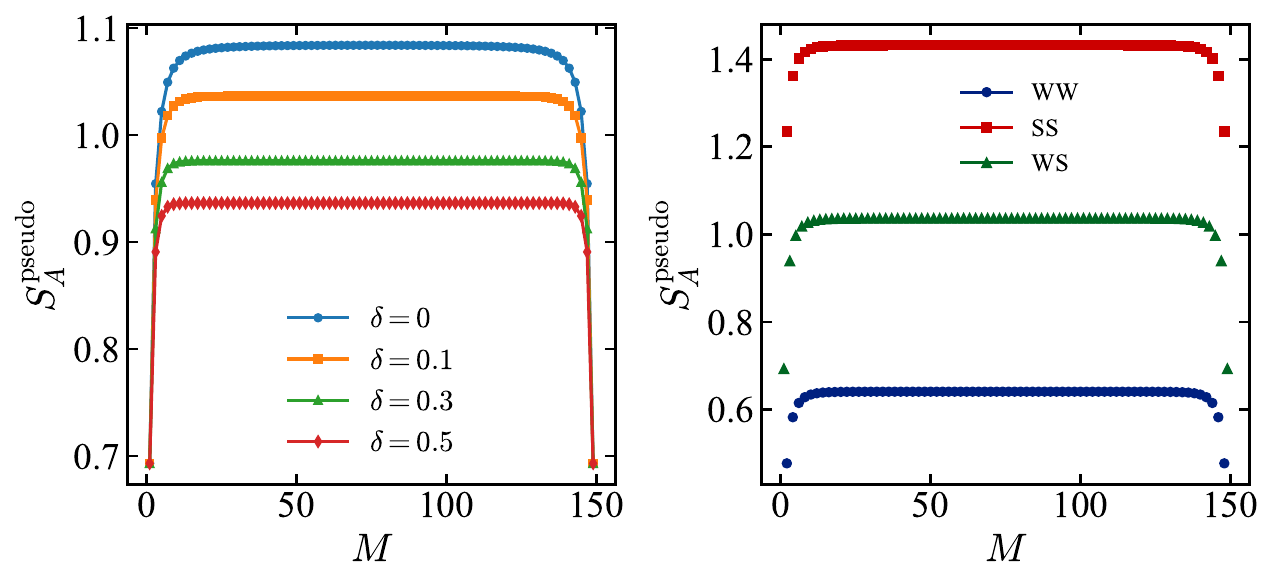}
        \caption{Trivial phase, $v_{\mathrm{ref}}=2.0>w=1$.}
        \label{fig: 2.0 PES}
    \end{subfigure}

    \caption{
    pseudo entropy $S_A^{\mathrm{pseudo}}$ as a function of subsystem size
    $M$ for SSH chain with $N=75$ unit cells and $w=1$ under periodic
    boundary conditions. In the right panels, WW, SS, and WS denote double
    cuts through weak-weak, strong-strong, and weak-strong bonds,
    respectively. (a) In the topological phase, $v_{\mathrm{ref}}=0.5$,
    the pseudo entropy saturates to an area-law value, with the WW cut
    giving the largest saturation value and the SS cut the smallest.
    (b) In the trivial phase, $v_{\mathrm{ref}}=2.0$, area-law saturation
    is also observed, but the cut hierarchy is reversed.
    }
    \label{fig:pes_saturation_gapped}
\end{figure*}

\begin{figure}[tbp]
    \centering

    \includegraphics[width=\columnwidth]
    {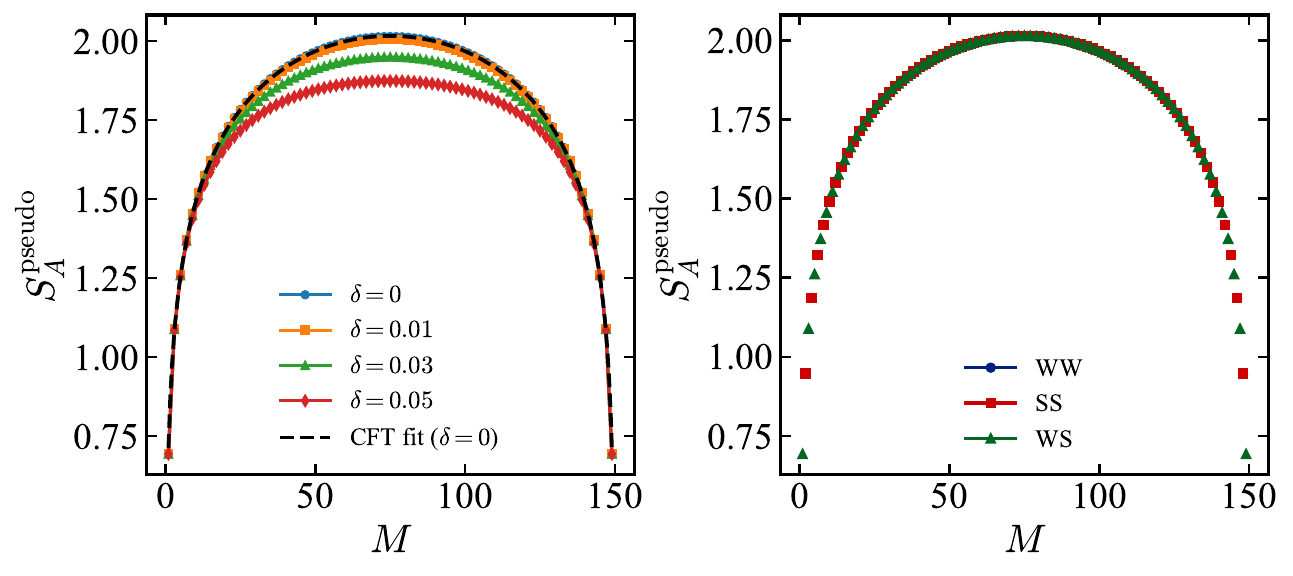}
    \caption{
    pseudo entropy $S_A^{\mathrm{pseudo}}$ at the critical point
    $v_{\mathrm{ref}}=1.0=w$.
    The pseudo entropy follows the logarithmic CFT scaling of
    equation~\eqref{eq:log scaling CFT}. The fitted coefficients
    $(a_{\mathrm{pseudo}},b_{\mathrm{pseudo}})$ are
    $(0.3338,0.7243)$, $(0.3301,0.7312)$,
    $(0.3056,0.7756)$, and $(0.2733,0.8326)$
    for $\delta=0,0.01,0.03,0.05$, respectively.
    At criticality, the distinction between WW, WS, and SS cuts disappears because the intercell and intracell hopping amplitudes are equal.
    }
    \label{fig: 1.0 PES}
\end{figure}

\section{Pseudo entropy of SSH model}
\label{sec:pseudo_SSH}
The Su--Schrieffer--Heeger (SSH) model, as depicted in figure \ref{fig:SSH_diagram}, is a one-dimensional tight-binding model used to describe spin-less electron hopping on two sites with alternating strengths \cite{Su:1979ua, Asboth2016}. The parameter $v$ denotes the intracell hopping amplitude between the two sites $A$ and $B$ inside the same unit cell, while $w$ denotes the intercell hopping amplitude between site $B$ of unit cell $m$ and site $A$ of unit cell $m+1$. For an open chain, the single-particle Hamiltonian is

\begin{align}
h
&=
v\sum_{m=1}^{N}
\left(
\ket{m,B}\bra{m,A}
+
\ket{m,A}\bra{m,B}
\right)
\nonumber\\
&\quad
+
w\sum_{m=1}^{N-1}
\left(
\ket{m+1,A}\bra{m,B}
+
\ket{m,B}\bra{m+1,A}
\right).
\label{eq:ssh_bra_ket_hamiltonian}
\end{align}
Consider a general single-particle state, $\ket{\psi}
=
\sum_{m=1}^{N}
\left(
a_m \ket{m,A}
+
b_m \ket{m,B}
\right)$,
where $a_m$ and $b_m$ are the probability amplitudes on sublattices $A$ and $B$ of unit cell $m$, respectively. Acting the Hamiltonian \eqref{eq:ssh_bra_ket_hamiltonian} on this state couples the two sublattices within the same unit cell, while $w$ couples neighboring unit cells.\\ 

The SSH Hamiltonian in second-quantized form for an open chain is:
\begin{align}
\hat H
&=
v\sum_{m=1}^{N}
\left(
c_{m,B}^{\dagger}c_{m,A}
+
c_{m,A}^{\dagger}c_{m,B}
\right)
\nonumber\\
&\quad
+
w\sum_{m=1}^{N-1}
\left(
c_{m+1,A}^{\dagger}c_{m,B}
+
c_{m,B}^{\dagger}c_{m+1,A}
\right).
\label{eq:ssh_second_quantized_hamiltonian}
\end{align}
where $\left( c_m,A \right)$ and $\left( c_m,B \right)$ denote two flavors of fermionic annihilation operators defined at lattice site $m$. Since the Hamiltonian is quadratic, it describes non-interacting spinless fermions.

For Periodic Boundary Condition (PBC), we connect the last unit cell to the first one, $c_{N+1,A}\equiv c_{1,A}$, and the Hamiltonian is: 
\begin{align}
\hat H_{\mathrm{PBC}}
=
v\sum_{m=1}^{N}
&\left(
c_{m,B}^{\dagger}c_{m,A}
+
\mathrm{h.c.}
\right) \nonumber \\
&+
w\sum_{m=1}^{N}
\left(
c_{m+1,A}^{\dagger}c_{m,B}
+
\mathrm{h.c.}
\right),
\label{eq:ssh_pbc_second_quantized}
\end{align}
where ``h.c.'' denotes the Hermitian conjugate. Due to the translation invariance of the periodic boundary condition, the Hamiltonian in the momentum basis decomposes into independent $2\times2$ momentum sectors $H=\bigoplus_{k} H(k)$, with each block given by
\begin{align}
H(k)&=
\begin{pmatrix}
0 & v + w e^{-ik} \\
v + w e^{ik} & 0
\end{pmatrix} \nonumber \\
&= d_x(k)\sigma_x
+
d_y(k)\sigma_y,
\label{eq: periodic}
\end{align}
where $\sigma_x$ and $\sigma_y$ are Pauli matrices, and $d_x(k)=v+w\cos k$, $d_y(k)=w\sin k.$ Diagonalizing $H(k)$ gives the two-band energy spectrum
\begin{equation}
E_{\pm}(k)=\pm\sqrt{v^{2}+w^{2}+2vw\cos k}.
\end{equation}
The positive and negative branches belong to the upper and lower energy bands, respectively, which close when $v=w$ at $k=\pi$. This is the signature of a topological phase transition. When $|v|>|w|$, we call it a trivial or non-topological phase, while when $|v|<|w|$, we call it a topological phase (topological insulator). The gap closes at $|v|=|w|$. This phase classification remains true for the Open Boundary Condition (OBC) as well, though the topological invariants are defined from the bulk.\\ 
To compute pseudo entropy in the SSH model, we consider two SSH Hamiltonians with the same $w$ but different $v_1$ and $v_2$, $    \hat H_1 \equiv \hat H(v_1,w)$ and $\hat H_2 \equiv \hat H(v_2,w)$. For $|v_1|<|w|$ and $|v_2|<|w|$, both states lie in the topological phase. For $|v_1|>|w|$ and $|v_2|>|w|$, both states lie in the trivial phase. Finally, when one of $|v_1|$ and $|v_2|$ is smaller than $|w|$ while the other is larger than $|w|$, the two states belong to different phases. Throughout our numerical analysis, we set $w=1$ unless noted explicitly.\\
Let us denote the ground states of $\hat H_1$ and $\hat H_2$ by $  \ket{\Psi_1} \equiv \ket{\Psi(v_1)}$ and $\ket{\Psi_2} \equiv \ket{\Psi(v_2)}$ respectively. Since the SSH Hamiltonian is quadratic, we can use the method from Section \ref{sec:Fermionic_correlator} to compute the pseudo entropy. The key ingredient is a weak-value correlation matrix:
\begin{equation}
    \langle c_i^\dagger c_j\rangle_{1|2}
    =
    \frac{
    \bra{\Psi_2}c_i^\dagger c_j\ket{\Psi_1}
    }{
    \braket{\Psi_2|\Psi_1}
    }.
\end{equation}
From these weak values, a pseudo-correlation matrix can be built; from its eigenvalues $\nu_k$, the pseudo entropy can be computed. For each pair, we can also compute the averaged excess pseudo entropy $ \Delta S_{12}$
\begin{equation}
    \Delta S_{12}
    =
    \mathrm{Re}\left[S(T_A^{1|2})\right]
    -
    \frac{1}{2}
    \left[
    S(\rho_A^{(1)})
    +
    S(\rho_A^{(2)})
    \right],
\end{equation}
where $ \rho_A^{(1)}
    =
    \mathrm{Tr}_B\ket{\Psi_1}\bra{\Psi_1}$ and $ \rho_A^{(2)}
    =
    \mathrm{Tr}_B\ket{\Psi_2}\bra{\Psi_2}$.\\

In figure~\ref{fig:Pseudo entropy for v=0.5}, we have plotted the pseudo entropy $S(T_A^{1|2})$ and the averaged
entanglement entropy
$\bigl[S(\rho_A^{(1)})+S(\rho_A^{(2)})\bigr]/2$ for the $N=60$ with open boundary conditions for the SSH Hamiltonian as functions of the $v_2$ but for different fixed values of $v_1$. So, the free parameters are $v_1$ and $v_2$, while $w$ is set to $1$. We observe that the pseudo entropy is, in general, different from the averaged entanglement entropy, and its profile changes noticeably as $v_2$ crosses the phase boundaries. A more detailed study of the phase-dependent behavior of pseudo entropy is deferred to Section \ref{sec:pseudo_conjecture_test}.\\

\subsection{Area law and Critical scaling}

Interestingly, for the SSH model, the real part of pseudo entropy shows scaling behavior analogous to entanglement entropy. It exhibits a similar area law, critical scaling, and bond-cut dependence. To study this, we compute pseudo entropy between $|\Psi_1(v_1)\rangle$ and $|\Psi_2(v_{\mathrm{ref}})\rangle$ for three different $v_{\text{ref}}$. These three $v_{\text{ref}}$ are supposed to represent the topological, critical, and trivial regimes. We further allow $v_1$ to vary with the parameter $\delta$ as $v_1=1+\delta$.\\
For the topological phase, setting $v_{\mathrm{ref}}=0.5<w=1$, the left panel of
figure~\ref{fig: 0.5 PES} shows that the pseudo entropy does exhibit
area-law-like saturation. The saturation plateau depends on the $\delta$ as the minimum gap is at  $ \Delta_{\mathrm{gap}}= 2|w-v_1|=2|\delta|.$ At $\delta = 0$, the correlation length diverges and pseudo entropy has the largest value. As $\delta$ increases, the system moves deeper into the gapped
topological phase with an exponentially decaying correlation function $C(r) \sim e^{-r/\xi},$ where $\xi$ is the correlation length. The pseudo entropy thus has a finite saturation value as $ S_A^{\mathrm{pseudo}}(M)
    =
    S_{\infty}^{\mathrm{pseudo}}
    +
    \mathcal{O}\!\left(e^{-M/\xi}\right),$ where $M$ is the subsystem size. The right panel of figure~\ref{fig: 0.5 PES} shows that for fixed $\delta$, the saturation value is dependent on the boundary cut, $  S_{\infty}^{\mathrm{pseudo}}(\mathrm{WW})
    >
    S_{\infty}^{\mathrm{pseudo}}(\mathrm{WS})
    >
    S_{\infty}^{\mathrm{pseudo}}(\mathrm{SS}),$ where WW, WS, and SS denote double-weak, mixed, and double-strong
cuts, respectively. A similar observation holds for the trivial phase, as shown in figure \ref{fig: 2.0 PES}. The hierarchy of the boundary cuts is opposite, though. 


For the critical phase, as shown in figure \ref{fig: 1.0 PES}, we observe that the area-law saturation is replaced by logarithmic scaling of the type
\begin{equation}
    S_A^{\mathrm{pseudo}}(M)
    \simeq
    a_{\mathrm{pseudo}}
    \log\left[
    \frac{L}{\pi}
    \sin\left(\frac{\pi M}{L}\right)
    \right]
    +b_{\mathrm{pseudo}},
    \label{eq:log scaling CFT}
\end{equation}
where $L$ is the total system size and $M$ is the subsystem size. The first term is similar to the behavior of entanglement entropy in a two-dimensional Conformal Field Theory with the central charge c = 1 \cite{Calabrese:2004eu,Holzhey_1994}, while the second term depends on the relevant parameters. At $\delta=0$, the fitted coefficient $a_{\mathrm{pseudo}}\approx 0.3338$ is consistent with $c/3$, as expected for the free-fermion critical point with central charge $c=1$. Furthermore, the right panel shows no cut dependency as for $v_1 = w$, the distinction between weak and strong bond disappears.

\begin{figure*}[tb]
    \centering

    \begin{subfigure}{0.49\textwidth}
        \centering
        \includegraphics[width=\linewidth]{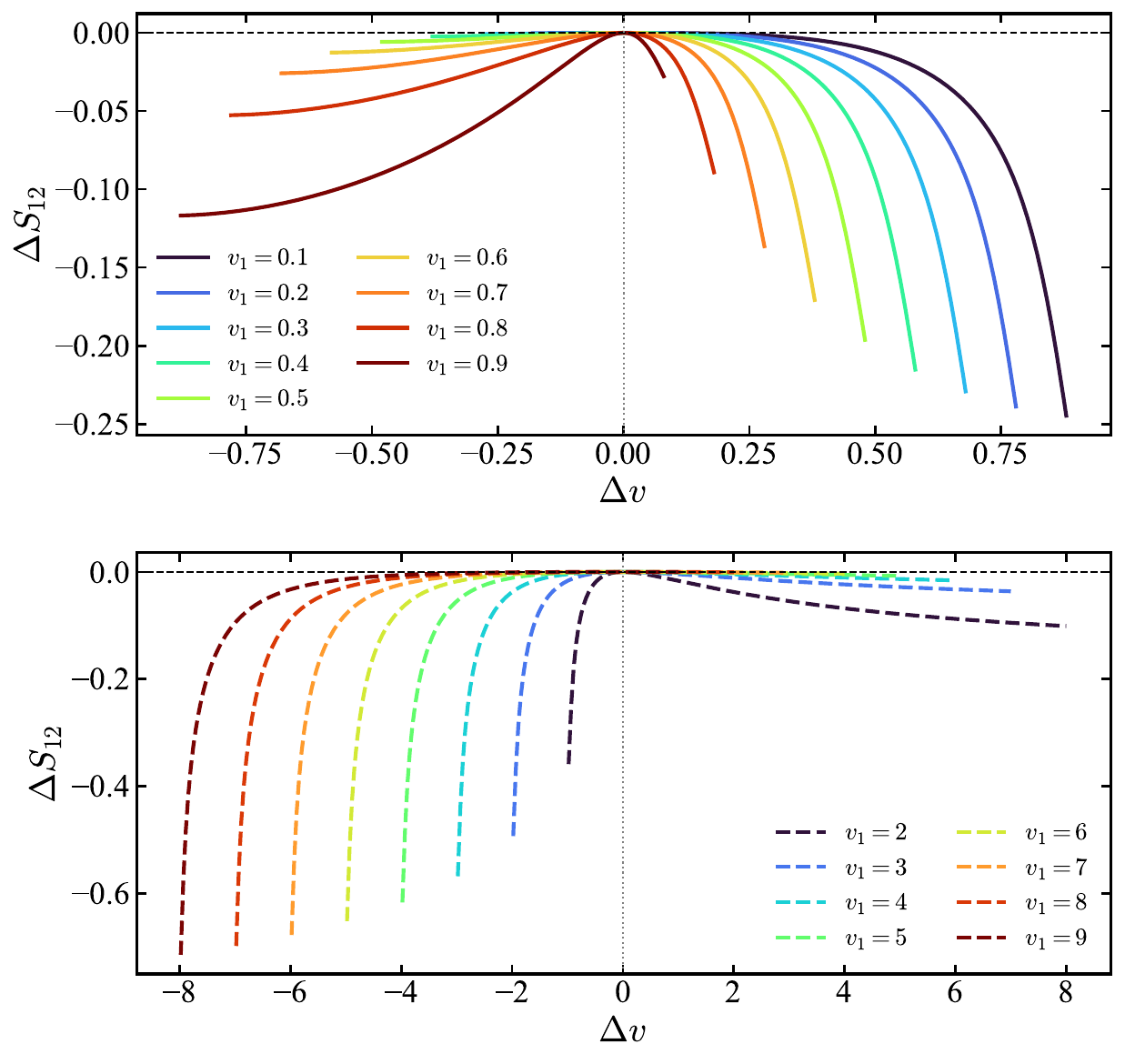}
        \caption{Same-phase transitions under PBC.}
        \label{fig: pbc same phase}
    \end{subfigure}
    \hfill
    \begin{subfigure}{0.49\textwidth}
        \centering
        \includegraphics[width=\linewidth]{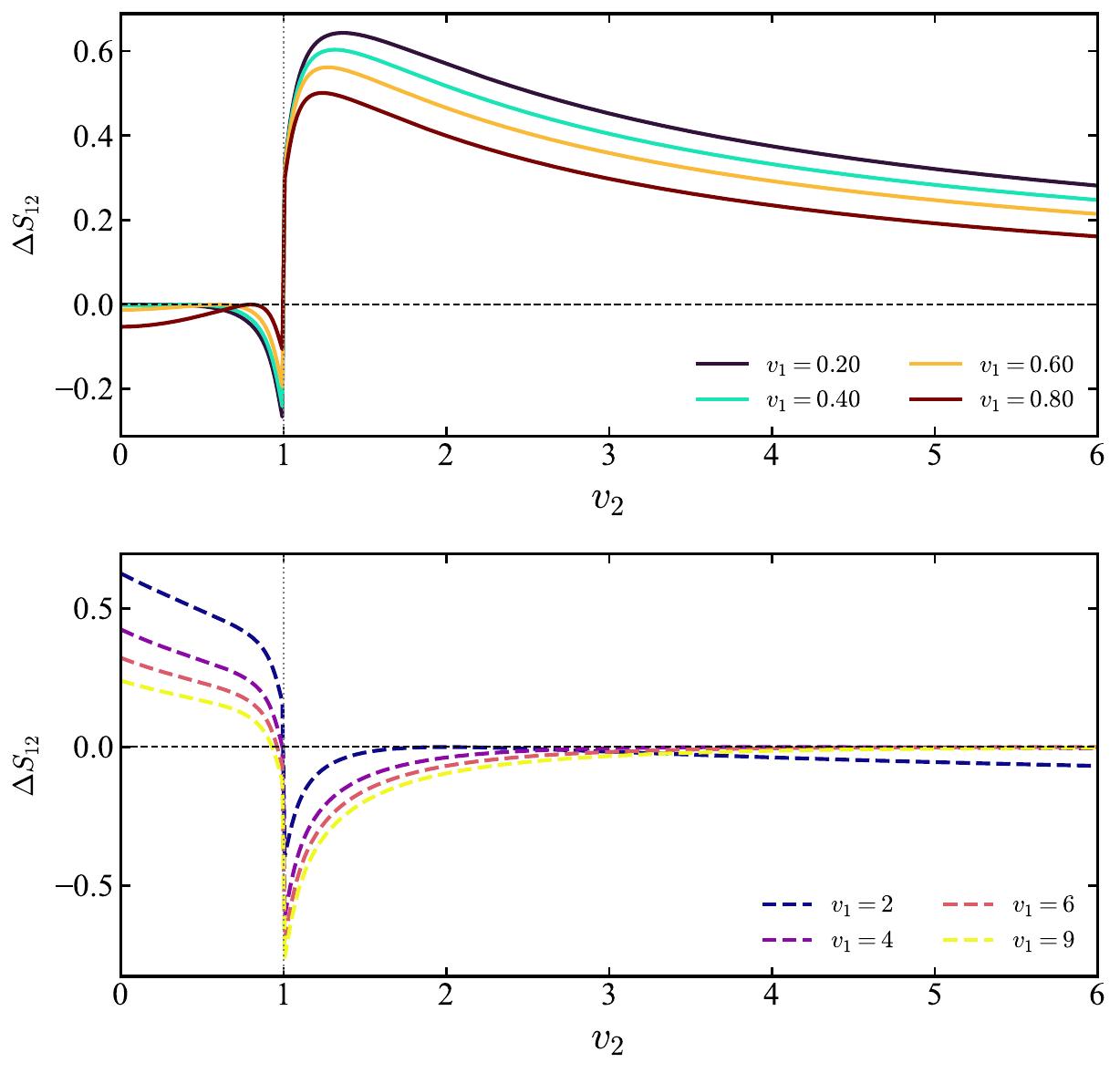}
        \caption{Cross-phase transitions under PBC.}
        \label{fig:pbc cross phase}
    \end{subfigure}

    \caption{pseudo entropy difference with average entanglement entropy $\Delta S_{12}$ under periodic boundary conditions. (a) Same-phase transitions with $N = 60$ and $w = 1$. The upper panel corresponds to transitions within the topological phase ($v_1 < w$), while the lower panel corresponds to transitions within the trivial phase ($v_1 > w$). (b) Cross-phase transitions with $N = 60$ and $w = 1$. The upper panel shows topological-to-trivial transitions ($v_1 < w$, $v_2 > w$), whereas the lower panel shows trivial-to-topological transitions ($v_1 > w$, $v_2 < w$).}
    \label{fig:pbc_results}
\end{figure*}

\begin{figure*}[tbp]
    \centering

    \begin{subfigure}{0.49\textwidth}
        \centering
        \includegraphics[width=\linewidth]{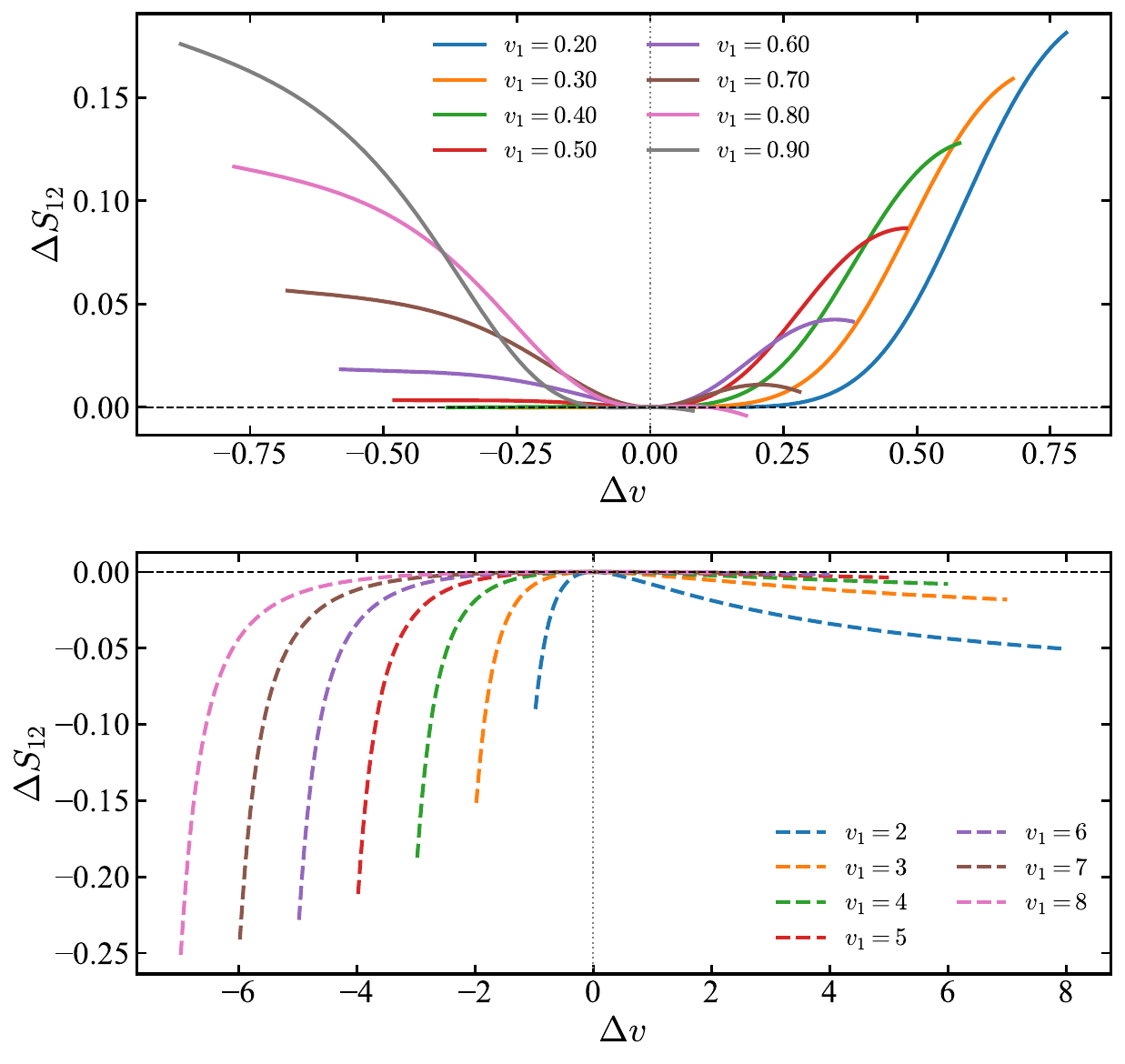}
        \caption{Same-phase transitions under OBC.}
        \label{fig:same phase}
    \end{subfigure}
    \hfill
    \begin{subfigure}{0.49\textwidth}
        \centering
        \includegraphics[width=\linewidth]{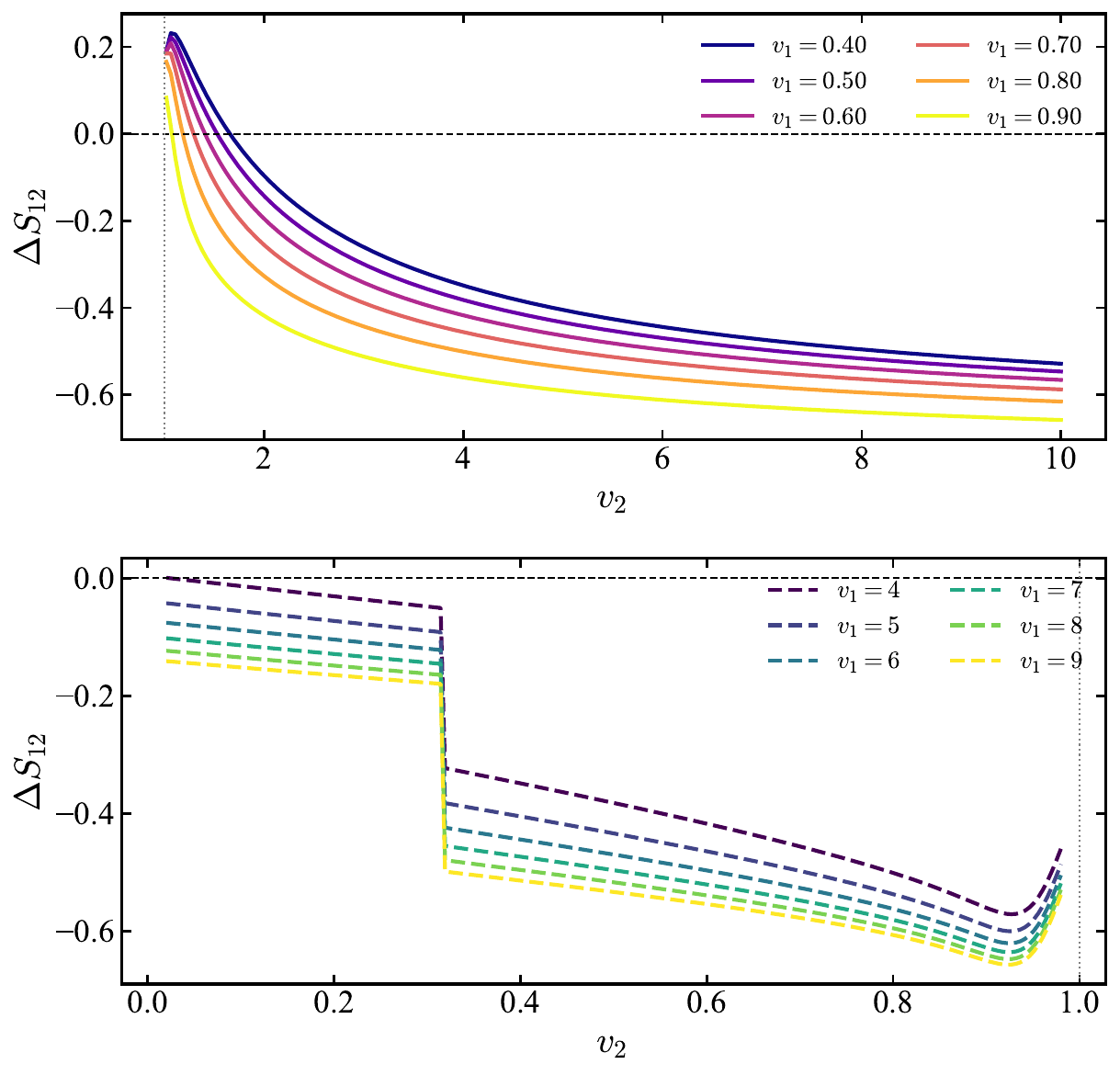}
        \caption{Cross-phase transitions under OBC.}
        \label{fig:cross phase}
    \end{subfigure}

    \caption{pseudo entropy difference with average entanglement entropy $\Delta S_{12}$ under open boundary conditions. (a) Same-phase transitions with $N=8$ and $w=1$. The upper panel corresponds to transitions within the topological phase ($v_1<w$), while the lower panel corresponds to transitions within the trivial phase ($v_1>w$). (b) Cross-phase transitions with $N=60$ and $w=1$. The upper panel shows topological-to-trivial transitions ($v_1<w$, $v_2>w$), whereas the lower panel shows trivial-to-topological transitions ($v_1>w$, $v_2<w$).}
    \label{fig:obc_results}
\end{figure*}


\section{Non-positivity test}
\label{sec:pseudo_conjecture_test}

Suppose, $S\!\left(\mathcal{T}^{\psi|\varphi}_A\right)$ is the pseudo entropy of the subsystem $A$ for the transition matrix between the quantum states $\ket{\psi}$ and $\ket{\varphi}$. The averaged excess entropy, $\Delta S \!\left(\mathcal{T}^{\psi|\varphi}_A\right)$, is defined as
\begin{align}
\Delta S \!\left(\mathcal{T}^{\psi|\varphi}_A\right) = &\,\text{Re} \!\left(S\!\left(\mathcal{T}^{\psi|\varphi}_A\right) \right) - \frac{1}{2} \left( S(\rho_\psi)_A +  S(\rho_\varphi)_A \right)
\label{eq: delta S}
\end{align}
where $ S(\rho_\psi)_A$ and $S(\rho_\varphi)_A$ are the entanglement entropies of the subsystem $A$ for the states $\ket{\psi}$ and $\ket{\varphi}$ respectively. 
In \cite{Mollabashi:2021xsd, Mollabashi:2020yie}, it was conjectured that the averaged excess entropy is non-positive,  
$ \Delta S \left(\mathcal{T}^{\psi|\varphi}_A\right) \leq 0$, when the states $\ket{\psi}$ and $\ket{\varphi}$ are in the same phase. However, if they are in different phases, $ \Delta S \left(\mathcal{T}^{\psi|\varphi}_A\right)$ tends to be positive, though it is not always necessary. This conjecture was argued based on the calculations in Lifshitz free scalar fields, holographic setups, and Ising and XY spin models \cite{Mollabashi:2021xsd, Mollabashi:2020yie}. In this section, we test the conjecture for the SSH model under both periodic and open boundary conditions. We find that while the conjecture holds for the periodic boundary condition, it doesn't hold for all system sizes $N$ under open boundary conditions. It appears to hold only in the large $N$ limit. In what follows, we will abbreviate $\Delta S \!\left(\mathcal{T}^{\psi|\varphi}_A\right)$ by $\Delta S_{12}$.\\ 
The numerical setup is as follows. We consider two SSH Hamiltonians \eqref{eq:ssh_pbc_second_quantized}
$\hat{H}(v_1,w)$ and $\hat{H}(v_2,w)$ with fixed $w=1$, and denote their
half-filled ground states by $|\Psi_1\rangle \equiv |\Psi(v_1)\rangle$ and
$|\Psi_2\rangle \equiv |\Psi(v_2)\rangle$. For each pair of parameters
$(v_1,v_2)$, we compute the corresponding pseudo and entanglement entropy. We then test the non-positivity prediction by varying $\Delta v=v_2-v_1$ both within a fixed phase and across the phase boundary at $v=w=1$.


We first test the non-positivity conjecture for the SSH chain under periodic
boundary conditions (PBC). Due to the translational symmetry in PBC, unlike an open chain, it doesn't support the localized edge states.  This allows us to test the bulk behavior of $\Delta S_{12}$.

Figure~\ref{fig: pbc same phase} shows the same-phase results for $N=60$ under PBC. In both the topological phase $(v_1,v_2<w)$ and the trivial phase
$(v_1,v_2>w)$,  $\Delta S_{12}$ remains non-positive throughout the range of $\Delta v=v_2-v_1$. As expected, the maximum value $\Delta S_{12} = 0$ occurs when the two states coincide at $\Delta v=0$. Therefore, the non-positivity test holds for the PBC. When two states are in different phases, as shown in figure~\ref{fig:pbc cross phase}, $\Delta S_{12}$ can be positive or negative, but it is still consistent with the conjecture. Interestingly, near the critical point $v_2=w=1$, the curve develops a sharp negative dip, reflecting the closing of the bulk energy gap. Overall, these
results show that, under PBC, $\Delta S_{12}$ provides a clear diagnostic of
whether the two SSH ground states belong to the same or different bulk phases.





 Unlike the periodic chain, the open SSH chain supports
boundary-localized edge modes in the topological phase. This allows us to test whether $\Delta S_{12}$ provides a probe of these boundary degrees of freedom. When states share the same phase but differ in hopping ratio, the edge modes contribute to $\Delta S_{\rho_A}$, inflating $\Delta S_{12}$ above zero for small $N$. This inflation vanishes as the localization length becomes negligible relative to the system size. 

\begin{figure}[tbp]
    \centering
    \includegraphics[width= 1 \linewidth]{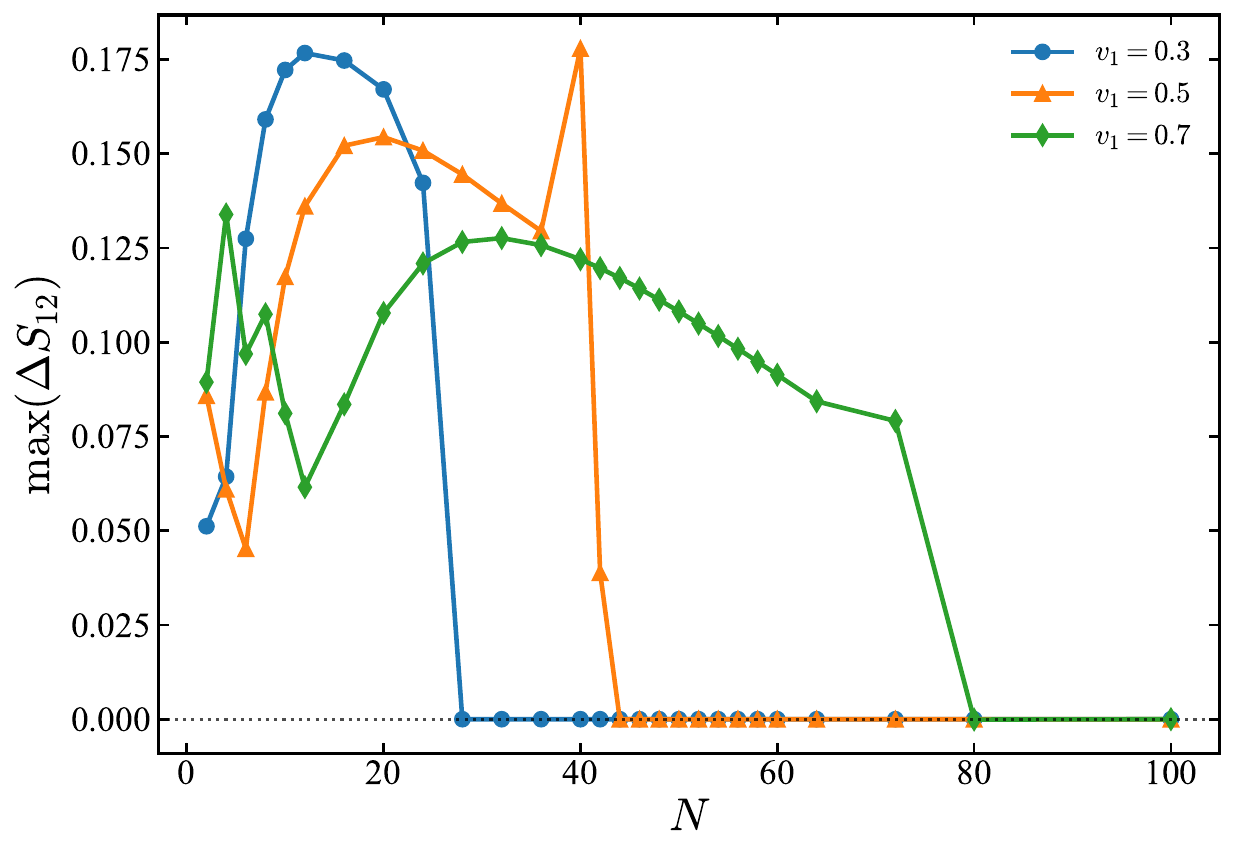}
    \caption{Finite-size scaling of the maximum pseudo entropy  difference $\max(\Delta S_{12})$ as a function of system size $N$. As $N$ increases $\max(\Delta S_{12})$ tends to be non-positive.}
    \label{fig:finite size scaling}
\end{figure}

\begin{figure}[tbp]
    \centering
    \includegraphics[width=1\linewidth]{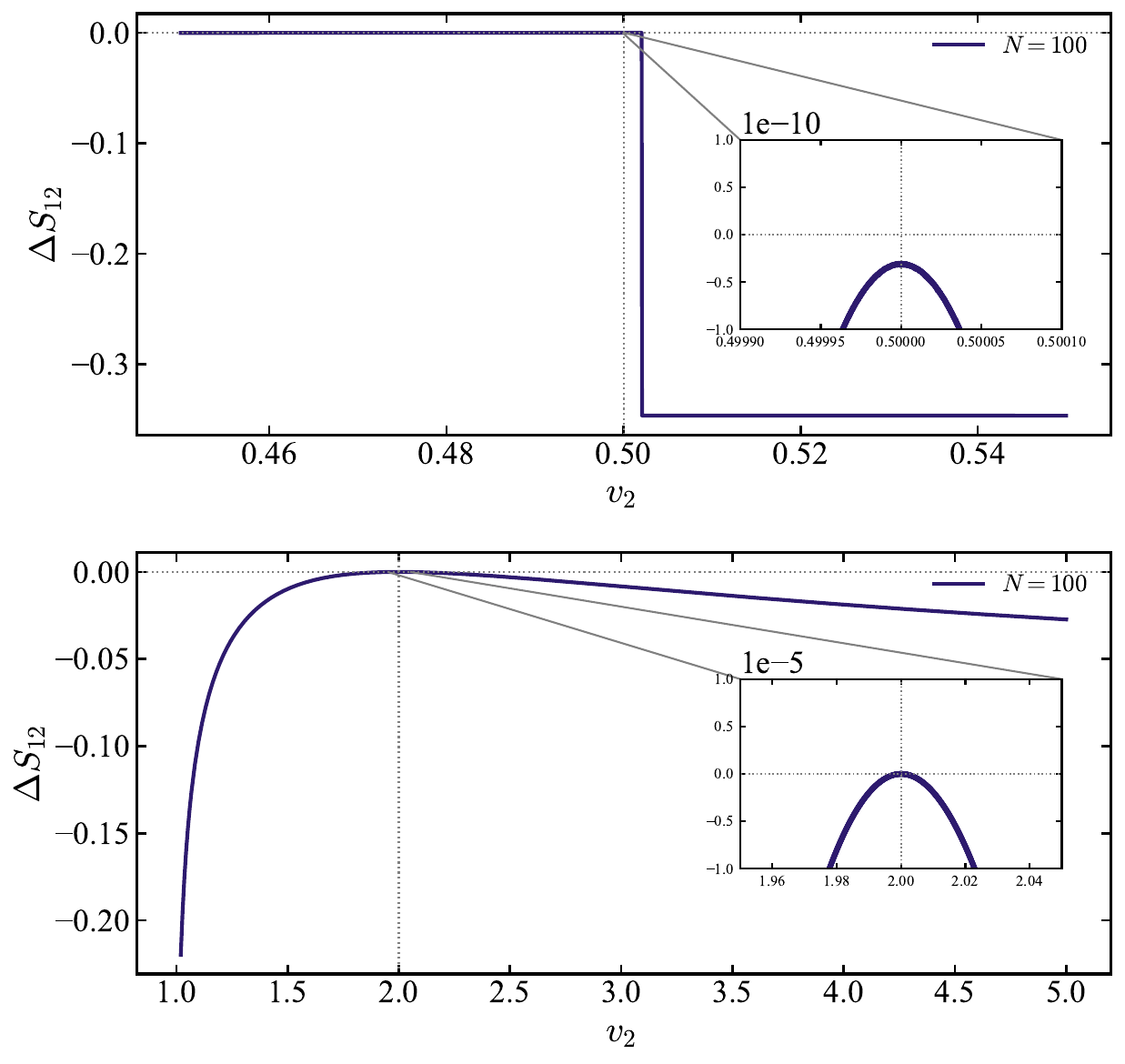}
    \caption{$\Delta S_{12}$ as a function of $v_2$ for same-phase transitions under OBC with $N=100$. The upper half plot is for the topological phase with $v_1=0.5$, and $w=1$, and the lower plot is for the trivial phase with $v_1=2$, and $w=1.$}
    \label{fig: For $n=60$}
\end{figure}

Figure~\ref{fig:same phase} shows the same-phase results for a small system size, $N=8$. The transitions within the topological phase ($v_1<w=1$), $\Delta S_{12}$ is positive over the entire range of $\Delta v$, violating the non-positivity as shown by the upper panel. In contrast, $\Delta S_{12}$ is consistently non-positive for the trivial phase ($v_1>w=1$), as shown in the lower panel. This suggests topological edge modes can strongly affect the averaged excess entropy for the finite-size scenario. In the case of $N=100$, $\Delta S_{12}$ is now non-positive throughout the range as shown in figure~\ref{fig: For $n=60$}. This supports the interpretation that the positive values observed for small $N$ are finite-size boundary effects rather than bulk violations of the conjecture. When the two ground states belong to different phases, $\Delta S_{12}$ can be either positive or negative depending on the Hamiltonian parameters, as shown in figure~\ref{fig:cross phase}. This is consistent with the conjecture, which requires only non-positivity for same-phase transitions and imposes no strict sign constraint on cross-phase comparisons. Figure \ref{fig:finite size scaling} shows that for large $N$, $\max(\Delta S_{12})$ tends to zero with $N$ depending on the parameter $v_1$.

Taken together, these results show that pseudo entropy remains sensitive to the topological phase structure of the open SSH chain, while also revealing finite-size boundary contributions
associated with edge modes.

\section{Imaginary Pseudo entropy  after a quench}
\label{sec:imS-DPT}

So far, we have considered only the real part of the pseudo entropy , since the transition matrix was constructed from two SSH ground states of a real-symmetric Hamiltonian. To get the imaginary contribution, we consider a quantum quench in the SSH chain. Interestingly, the imaginary part provides a complementary diagnosis of dynamical phase transitions (DPTs). To define DPT, we have to first consider the dynamical free-energy density $r(t) =-\frac{1}{N}\log |G(t)|^2$, where $G(t)=\bra{\psi_0}e^{-iH_1t}\ket{\psi_0}$ is the Loschmidt amplitude,  $\ket{\psi_0}$ the ground state of the pre-quench Hamiltonian $H_0$, and the quench dynamics is via $H_1\neq H_0$. A DPT is then identified by a non-analyticity of $r(t)$ at a critical time $t_n^*$ \cite{LeClair:1995uf, Heyl:2013ywe, Heyl:2017blm}. The relation between DPT and the topology of quench is sharpened due to the theorem by Vajna and D\'ora \cite{PhysRevB.91.155127}. The SSH Bloch Hamiltonian is
 $H_\alpha(k) = \mathbf{d}_\alpha(k)\!\cdot\!\bm{\sigma}$
where $\alpha=0,1$ labels the pre- and post-quench Hamiltonians. The many-body Loschmidt amplitude factorizes over momenta as
\begin{equation}
G(t)=\prod_{k\in\BZ}\Big[
\cos(\varepsilon_k^1 t)
+i\,\hat{\mathbf d}_0(k)\!\cdot\!\hat{\mathbf d}_1(k)
\sin(\varepsilon_k^1 t)
\Big],
\label{eq:G-product}
\end{equation}
with $\varepsilon_k^1=|\mathbf d_1(k)|$. A dynamical phase transition occurs when there exists a critical momentum $k^*$ satisfying
\begin{equation}
\hat{\mathbf d}_0(k^*)\!\cdot\!\hat{\mathbf d}_1(k^*)=0,
\quad
t_n^*=\frac{(n+\tfrac12)\pi}{\varepsilon_{k^*}^1},
\quad n\in\mathbb Z_{\geq0}.
\label{eq:VD-condition}
\end{equation}
For the SSH chain, such a critical momentum is guaranteed when the quench crosses the topological phase boundary\cite{PhysRevB.91.155127, Su:1979ua,Asb_th_2016}. Therefore, topology-changing quenches necessarily produce DPTs, whereas quenches within the same phase do not generically do so.


 We now construct the Slater-determinant pseudo-correlation matrix for the quench. Let $\Psi_0\in\mathbb R^{2N\times N}$ denote the matrix whose columns are the occupied single-particle orbitals of the half-filled ground state of $H_0$. Under the post-quench evolution, the occupied orbitals become
\begin{equation}
\Phi(t)=U_1(t)\Psi_0,
\qquad
U_1(t)=e^{-iH_1t}.
\label{eq:Phi-t}
\end{equation}
The corresponding pseudo‑correlation matrix is
\begin{equation}
\widetilde C(t)
=
\Phi(t) M(t)^{-1}\Psi_0^\dagger,
\qquad
M(t)=\Psi_0^\dagger\Phi(t) .
\label{eq:C-quench}
\end{equation}
Restricting to the subsystem $A$ gives $\widetilde C_A(t)=P_A\widetilde C(t)P_A .$ Here $\widetilde{C}(t)$ is the time-dependent version of $C^{1|2}$, 
specialised to the quench setting.
The pseudo entropy is then obtained from the eigenvalues of $\widetilde C_A(t)$ using the free-fermion Peschel formula \cite{Peschel:2002yqj, Mollabashi:2021xsd, Murciano:2022lsw}. Since $\Phi(t)$ is generically complex, $\widetilde C_A(t)$ is non-Hermitian, and its spectrum can produce a nonzero imaginary part of the pseudo entropy.

The connection to the Loschmidt diagnostic is already evident in the overlap matrix.  The overlap matrix entering the pseudo-correlation matrix becomes singular whenever the Loschmidt amplitude vanishes, suggesting a direct connection between the imaginary pseudo entropy and dynamical phase transitions \cite{Misumi:2026jha}. A detailed analysis is left for future work. Here, we will restrict ourselves to the numerical simulations.

\begin{figure}[tbp]
    \centering
    \includegraphics[width=1\linewidth]{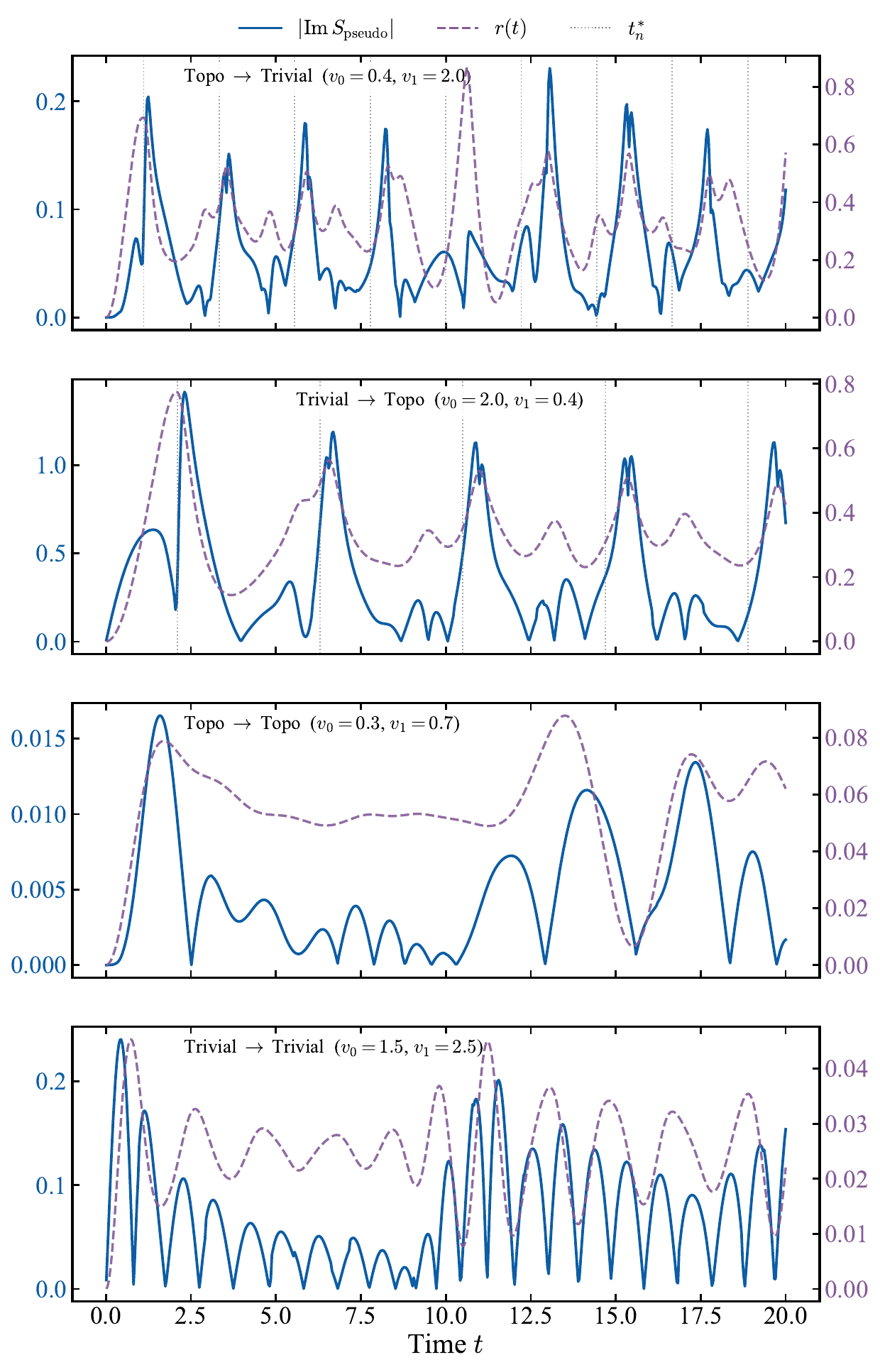}
    \caption{Test of the imaginary-pseudo entropy as a DPT diagnostic for four representative SSH quenches at $w=1$, $N=20$ under periodic boundary conditions at half filling. Top two panels: topology-crossing quenches ($|\Delta\nu|=1$) show sharp, quasi-periodic peaks of $\Im \Spseudo$ (blue, left axis) coincident with cusps of the Loschmidt rate $r(t)$ (purple, right axis), at the critical times $t_n^* = (n+\frac{1}{2})\pi/\epsilon_{k^*}^1$ predicted by equation~\eqref{eq:VD-condition} and indicated by dashed lines. Bottom two panels: within-phase quenches ($\Delta\nu=0$) admit no real solution $k^*$ of the orthogonality condition; $r(t)$ correspondingly stays an order of magnitude smaller (note rescaled axes), and the residual structure in $\Im \Spseudo$ is non-periodic and uncorrelated with any predicted DPT time.}
    \label{fig:fourquench}
\end{figure}

\begin{figure}[tbp]
    \centering
    \includegraphics[width= \linewidth]{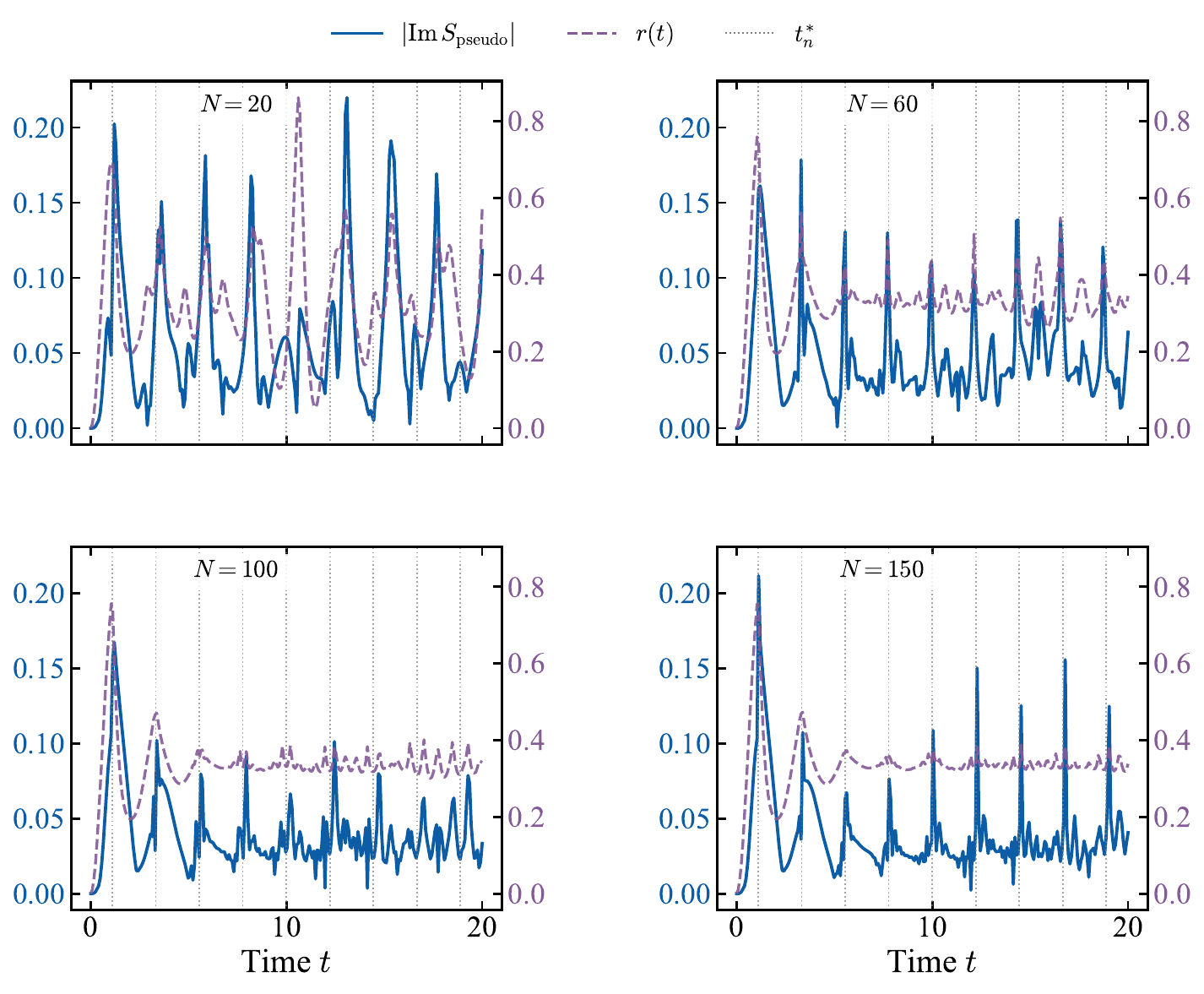}
    \caption{Half-chain von Neumann pseudo entropy  $|\mathrm{Im}\,S_{\mathrm{pseudo}}|$ (blue, left axis) and Loschmidt rate $r(t)$ (orange, right axis) for the topology-crossing quench $v_{0}=0.4 \to v_{1}=2.0$ at $w=1$, shown for system sizes $N=20,\,60,\,100,\,150$ ($2N$ sites, periodic boundary conditions). Vertical dashed lines mark the Vajna--D\'ora critical times $t_n^* = (n+\tfrac{1}{2})\pi/\varepsilon_{k^*}^1$ at which the Loschmidt amplitude develops Fisher zeros. The peaks of $|\mathrm{Im}\,S_{\mathrm{pseudo}}|$ and the cusps of $r(t)$ both align with
$t_n^*$ at every $N$; the $r(t)$ cusps sharpen as $N$ increases, consistent with approach to the thermodynamic-limit non-analyticity.}
    \label{fig:scalability}
\end{figure}

 Figure~\ref{fig:fourquench} compares the magnitude of the imaginary part of the pseudo entropy, $|\operatorname{Im}S_{\mathrm{pseudo}}|$, with the dynamical free-energy density $r(t)$ for four representative SSH quenches under periodic boundary conditions. We use $|\mathrm{Im}\,S_{\mathrm{pseudo}}|$ since exchanging the two boundary states conjugates $S_{\mathrm{pseudo}}$, so the sign of its imaginary part is fixed only by the arbitrary choice of reference state. The magnitude is invariant under this choice and preserves the diagnostic enhancement at the critical times $t_n^{*}$. When the quenches cross topological phases, we observe that peaks of $|\operatorname{Im}S_{\mathrm{pseudo}}|$ align with the cusps of $r(t)$ and with the critical times predicted by equation~\eqref{eq:VD-condition}. In contrast, for same-phase quenches, $|\operatorname{Im}S_{\mathrm{pseudo}}|$ is not correlated with DPT times. This shows that $\operatorname{Im}S_{\mathrm{pseudo}}$ does provide a sensitive diagnosis of DPTs in the SSH quench. Furthermore, figure~\ref{fig:scalability} shows that the peaks of $|\operatorname{Im}S_{\mathrm{pseudo}}|$ become sharper with system size. This provides evidence that pseudo entropy acts as an order-parameter-like probe of dynamical criticality. Compared with other measures, such as fidelity susceptibility and the Loschmidt rate, pseudo entropy provides a local-temporal measure that can yield a finer diagnosis of DPT. We plan to explore the relationship between pseudo entropy  and DPT in future work.

\section{Conclusion and Outlook}
\label{sec:outlook}
In this work, we studied pseudo entropy and the averaged excess entropy $\Delta S_{12}$ within the Su--Schrieffer--Heeger (SSH) model to characterize topological phase transitions. A central question was whether $\Delta S_{12}$ respects the conjectured sign structure: non-positive within the same phase and not-sign definite in phase crossing. Under periodic boundary conditions, the conjecture is fully satisfied with no exceptions at all system sizes and across all three phases: topological, trivial, and crossing. However, under open boundary conditions, this conjecture holds only for sufficiently large system sizes, indicating that the non-positivity is a feature of the thermodynamic limit rather than a finite-size property. We also showed that pseudo entropy  obeys a one-dimensional area law, which breaks down at the critical point and follows logarithmic CFT scaling. Beyond the equilibrium analysis, we showed that an imaginary component arises in the pseudo entropy  upon a sudden quench of the SSH chain. $|\mathrm{Im}\, S_{\mathrm{pseudo}}|$ acts as a diagnostic for dynamical phase transitions: its singular peaks coincide precisely with the Fisher zeros of the Loschmidt amplitude, and the peaks sharpen with system size, approaching genuine non-analyticities in the thermodynamic limit~\cite{Heyl:2013ywe, Heyl:2017blm, PhysRevB.91.155127}.

Several interesting future directions remain open. Some of them are as follows. The most immediate is the extension to the non-Hermitian setting~\cite{Lieu:2018cbt, Fu:2025fle,Guo:2023tjv}, where we can test whether pseudo entropy tracks phase transitions once unitarity is lost. Second, a natural next step from the simple SSH model is to a 1-dimensional p-wave superconductor - the Kitaev chain, which also displays the topological phase transitions~\cite{Caputa:2022yju}. Third, in analogy with the conventional entanglement spectrum, one can also explore the full pseudo-entanglement spectrum $\{\nu_k\}$ which may encode subtle topological information \cite{Nishioka:2021cxe}. So far, our model is primarily based on free theory, where we rely on the correlation matrix method for numerical calculations. This suggests an extension to non-integrable, interacting models~\cite{Karrasch:2013nis, Jurcevic2017}, where the correlation-matrix methods no longer apply and tensor-network or exact-diagonalization techniques become necessary. Finally, the relation with other measures, such as complexity, also needs to be analyzed \cite{Adhikari:2025vdl, Adhikari_2024}.

\acknowledgments

This research is part of the Abdus Salam International Centre for Theoretical Physics (ICTP) program: Physics Without Frontiers (PWF) and we acknowledge support from the PWF program of the ICTP, Italy.
The research is part of the Munich Quantum Valley, which is supported by the Bavarian state government with funds from the Hightech Agenda Bayern Plus.

\bibliographystyle{apsrev4-2}
\bibliography{ref1.bib}

\clearpage
\onecolumngrid

\appendix
\section{Correlation matrices of Fermionic systems and Pseudo Entropy}
\label{apx:fermionic_pseudo}
In this section we calculate the pseudo entropy of fermionic systems using the correlation matrix method. We begin by defining two free fermion states and the weak value of the transition matrix. Exploiting the Gaussian structure of the reduced transition matrix, we encode all two-point weak values into a single $2N_A\times 2N_A$ matrix $i\Gamma$, whose eigenvalues $\pm\nu_k$ directly determine the pseudo entropy.
\subsection{Two Free-Fermion States and the Weak Value Transition Matrix}
\label{subsec:setup}
We consider a lattice with $N$ sites. At each site $i$, we have a fermionic mode described by the annihilation operator $c_i$ and the creation operator $\cdag_i$. These operators satisfy the canonical anti-commutation relations $\{c_i,\, \cdag_j\} \;=\; \delta_{ij}$. We consider two generic quadratic Hamiltonians of a system of free fermions hopping between lattice sites.

\begin{equation}
  H^{(s)} \;=\; \sum_{i,j=1}^{N} h^{(s)}_{ij}\, \cdag_i c_j, \qquad s = 1, 2
  \label{eq:two_hamiltonians}
\end{equation}
Each $h^{(s)}$ is a Hermitian $N\times N$ single-particle matrix. For $s = 1, 2$, we diagonalise $h^{(s)}$ with a unitary matrix $U^{(s)}$:
\begin{equation}
  \left(U^{(s)}\right)^\dagger h^{(s)}\, U^{(s)}
  \;=\;
  \mathrm{diag}\!\left(\epsilon_1^{(s)}, \epsilon_2^{(s)}, \ldots, \epsilon_N^{(s)}\right).
  \label{eq:diag_each}
\end{equation}
This defines the normal modes for each Hamiltonian:
\begin{equation}
  \eta_k^{(s)} \;=\; \sum_i \left(U^{(s)}_{ik}\right)^*\! c_i,
  \qquad
  \left(\eta_k^{(s)}\right)^\dagger \;=\; \sum_i U^{(s)}_{ik}\, \cdag_i
  \label{eq:normal_modes_s}
\end{equation}
In the normal-mode basis, the Hamiltonian is diagonal as
$H^{(s)} \;=\; \sum_k \epsilon_k^{(s)}\, \eta_k^{(s)\dagger} \eta_k^{(s)}$. Filling the $M_s$ lowest modes, the two ground states are:
\begin{equation}
  |\psi_s\rangle = \prod_{k=1}^{M_s}(\eta^{(s)}_k)^\dagger |0\rangle
  \label{eq:ground_states}
\end{equation}
Both $|\psi_1\rangle$ and $|\psi_2\rangle$ are Slater determinants, i.e.\ Gaussian states for which every multi-point expectation value factorises into products of two-point functions via Wick's theorem. We split the system into subsystem $A$ with $\NA$ sites (the region of interest) and its complement $B$ with $N-\NA$ sites. Now, we consider the post-selected generalization of the expectation value of an operator, called weak value. This reduces the computation of the pseudo entropy to computing weak values of quadratic operators on subsystem $A$. For any operator $O$, it is defined as:

\begin{equation}
  \wv{O} \;\equiv\;
  \frac{\langle\psi_2|O|\psi_1\rangle}{\langle\psi_2|\psi_1\rangle}
  \;=\; \mathrm{Tr}\!\left[O\,\tauFull\right].
  \label{eq:weak_value}
\end{equation}
The second equality can be verified directly:
\begin{align}
  \mathrm{Tr}\!\left[O\,\tauFull\right]
  &= \frac{1}{\langle\psi_2|\psi_1\rangle}\,
     \mathrm{Tr}\!\left[O\,|\psi_1\rangle\langle\psi_2|\right]
   = \frac{1}{\langle\psi_2|\psi_1\rangle}
     \sum_n \langle n|O|\psi_1\rangle\langle\psi_2|n\rangle
   = \frac{\langle\psi_2|O|\psi_1\rangle}{\langle\psi_2|\psi_1\rangle}
   = \wv{O}.
  \label{eq:wv_proof}
\end{align}
When $O$ is localised in subsystem $A$, this becomes
$\wv{O}=\mathrm{Tr}_A[O\,\tauA]$, so the weak value directly probes
the reduced transition matrix.

\subsection{Gaussian Structure of the Reduced Transition Matrix}
\label{subsec:gaussian}

Since both $|\Psi_1\rangle$ and $|\Psi_2\rangle$
are Gaussian (Slater determinant) states, the operator $|\Psi_1\rangle\langle\Psi_2|$ in equation \eqref{eq:transition}
is a Gaussian operator in Fock space. Taking the partial trace $\mathrm{Tr}_B$ of a Gaussian operator over part of the Hilbert space gives another Gaussian operator (this is the fermionic analogue of marginalising a Gaussian probability distribution, which again gives a Gaussian). Therefore, $\tauA$ must have the form:

\begin{equation}
\tauA \;=\; \frac{1}{\mathcal{Z}}\, e^{-\mK},
\qquad
\mK \;=\; \sum_{i,j\in A} K_{ij}\, \cdag_i c_j
\label{eq:TA_gaussian}
\end{equation}

where $K$ is an $\NA\times\NA$ matrix and
$\mathcal{Z}=\mathrm{Tr}_A[e^{-\mK}]$ is the normalisation. Each single-mode thermal state is completely characterized by its occupation number, and consequently $\rho$ is entirely determined by the full collection of two-point expectation values. Reverting to the original operators via $c_i = \sum_k U_{ik}\, \eta_k$, every fermionic Gaussian state (f.g.s.) is completely characterized by the correlators:
\begin{equation}
  C_{ij}^{\cdag c} := \langle\cdag_i c_j\rangle,
  \qquad
  C_{ij}^{cc} := \langle c_i c_j\rangle.
  \label{eq:two_correlators}
\end{equation}
We collect these correlators into the so-called correlation
matrix

\begin{equation}
    C := \langle\cdag c\rangle
  = \begin{pmatrix}
      C^{\cdag c} & C^{\cdag\cdag} \\[4pt]
      C^{cc}    & C^{c\cdag}
    \end{pmatrix},
  \label{eq:C_full}
\end{equation}

with
\begin{align}
  C_{ij}^{\cdag c}   &= \langle\cdag_i c_j\rangle,
  \label{eq:Ccc_def}\\[4pt]
  C_{ij}^{cc}      &= \langle c_i c_j\rangle,
  \label{eq:Caac_def}\\[4pt]
  C_{ij}^{\cdag\cdag}  &= \langle\cdag_i \cdag_j\rangle
                    = -\overline{C_{ij}^{cc}},
  \label{eq:Caadag_def}\\[4pt]
  C_{ij}^{c\cdag}    &= \langle c_i \cdag_j\rangle
                    = \bigl(\mathbf{1} - C^{\cdag c}\bigr)_{ij}.
  \label{eq:Cdagc_def}
\end{align}

The last two relations follow from the canonical anti-commutation relations: $\langle\cdag_i\cdag_j\rangle = -\overline{\langle c_i c_j\rangle}$
(from $\{\cdag_i,\cdag_j\}=0$) and
$\langle c_i\cdag_j\rangle = \delta_{ij} - \langle\cdag_j c_i\rangle$
(from $\{c_i,\cdag_j\} = \delta_{ij}$).
The Hamiltonian \eqref{eq:two_hamiltonians} conserves particle number and consequently the anomalous correlators vanish identically, $C_{ij}^{cc}=\langle c_i c_j\rangle=0$, and the full correlation matrix \eqref{eq:C_full} reduces to the single block
\begin{equation}
  C_{ij} \equiv C_{ij}^{\dagger c} = \langle c_i^{\dagger}c_j\rangle.
  \label{eq:C_simple}
\end{equation}
This $N\times N$ Hermitian matrix $C$ is the central object used throughout the rest of this work. we employ the right eigen-decomposition of the non-Hermitian matrix, for which there exists an invertible matrix $W$ such that:

\begin{equation}
  K \;=\; W\, \mathrm{diag}\!\left(\varepsilon_1, \varepsilon_2, \ldots,\varepsilon_\ell\right) W^{-1}
  \label{eq:K_eigen}
\end{equation}

where $\varepsilon_k$ are the (generally complex) eigenvalues of $K$. Thus in new modes $\mK \;=\; \sum_k \varepsilon_k\, d_k^\dagger d_k$ and the exponential factorises over independent modes:
\begin{equation}
  e^{-\mK} \;=\; \prod_{k=1}^{\ell} e^{-\varepsilon_k\, d_k^\dagger d_k}
  \label{eq:exp_factorize}
\end{equation}
Since each mode has only two states, $\Ket{0}_k$ and $\Ket{1}_k$, corresponding to the occupation eigenvalues of $d_k^\dagger d_k$,

\begin{align*}
  \Tr\!\left[e^{-\varepsilon_k d_k^\dagger d_k}\right]
  &= \prescript{}{k}{\bra{0}}\, e^{-\varepsilon_k \cdot 0}\, \Ket{0}_k
   + \prescript{}{k}{\bra{1}}\, e^{-\varepsilon_k \cdot 1}\, \Ket{1}_k \\
  &= 1 + e^{-\varepsilon_k}
\end{align*}

Over all $\NA$ modes, the normalizing factor $\mathcal{Z}$ becomes:
\begin{equation}
 \mathcal{Z} = \prod_{k=1}^{\NA} \left(1 + e^{-\varepsilon_k}\right)
\end{equation}
Defining the normal-mode number operators $n_{A,k}$ (whose eigenvalues are $0$ or $1$), the diagonal form of
$\tauA$ is:
\begin{equation}
\tauA
=
\bigotimes_{k=1}^{\NA}
\frac{
e^{-\varepsilon_k\, n_{A,k}}
}{
1 + e^{-\varepsilon_k}
}
\label{eq:tau_diagonal}
\end{equation}
The task now reduces to finding the $\NA$ complex eigenvalues $\{\epsilon_k\}$.  We do this by matching the weak values of quadratic operators computed from equation \eqref{eq:tau_diagonal} to the same weak values computed directly from $|\psi_1\rangle$ and $|\psi_2\rangle$.  This
matching is most elegantly organised using the Majorana representation, which we introduce next. 

\subsection{The Majorana Representation}
\label{subsec:majorana}

The weak values of all quadratic operators on $A$ can be encoded in a single $2\NA\times 2\NA$ matrix.  The cleanest way to do this for fermions is to introduce the Majorana operators, which are the fermionic analogues of the scalar field $\phi$ and momentum $\pi$ in the bosonic case.  For each site $m\in A$, we define:
\begin{equation}
  \tilde{c}_{2m} \;=\; \cdag_m + c_m,
  \qquad
  \tilde{c}_{2m-1} \;=\; i\!\left(\cdag_m - c_m\right).
  \label{eq:majorana}
\end{equation}
These operators are Hermitian ($\tilde{c}_m^\dagger = \tilde{c}_m$) and
satisfy the Majorana anti-commutation relation:
\begin{equation}
  \{\tilde{c}_m,\,\tilde{c}_n\} = 2\delta_{mn}.
  \label{eq:majorana_CAR}
\end{equation}
Each original site $m$ contributes two Majorana operators
$\tilde{c}_{2m}$ and $\tilde{c}_{2m-1}$, so $\NA$ sites give $2\NA$ Majorana operators in total exactly the right size for the $2\NA\times 2\NA$ matrix we need. Inverting equation \eqref{eq:majorana}, the original operators are recovered as:
\begin{align}
  \tilde{c}_{2m} - i\tilde{c}_{2m-1}
  &= (\cdag_m + c_m) - i\cdot i(\cdag_m - c_m)
   = \cdag_m + c_m + \cdag_m - c_m = 2\cdag_m, \notag\\[4pt]
  \tilde{c}_{2m} + i\tilde{c}_{2m-1}
  &= (\cdag_m + c_m) + i\cdot i(\cdag_m - c_m)
   = \cdag_m + c_m - \cdag_m + c_m = 2c_m,
  \label{eq:c_from_majorana_derivation}
\end{align}
which gives:
\begin{equation}
  \cdag_m = \frac{\tilde{c}_{2m} - i\tilde{c}_{2m-1}}{2},
  \qquad
  c_m = \frac{\tilde{c}_{2m} + i\tilde{c}_{2m-1}}{2}.
  \label{eq:c_from_majorana}
\end{equation}

\subsection{The Four Weak-Value Correlators and the \texorpdfstring{$\Gamma$} {Matrix} Matrix }

Using the weak value defined in \eqref{eq:weak_value}, we define the four $\NA\times\NA$ matrices of two-point weak values:
\begin{equation}
  \mathcal{W}_1 \equiv \wv{c_m c_n},
  \quad
  \mathcal{W}_2 \equiv \wv{c_m \cdag_n},
  \quad
  \mathcal{W}_3 \equiv \wv{\cdag_m c_n},
  \quad
  \mathcal{W}_4 \equiv \wv{\cdag_m\cdag_n},
  \qquad m,n\in A.
  \label{eq:C1234_def}
\end{equation}
These are generalisations of the ordinary equal-time correlation matrix $\langle\cdag_m c_n\rangle$ to the two-state setting.  In particular, $\mathcal{W}_3$ plays the role of the ordinary correlation matrix, while $\mathcal{W}_1$ and $\mathcal{W}_4$ are anomalous (pairing) correlators that vanish when particle number is conserved. We now compute $\wv{\tilde{c}_m\tilde{c}_n}$ for all pairs $m,n\in A$ using equations \eqref{eq:C1234_def}.  Substituting the Majorana definitions \eqref{eq:majorana}, each of the four blocks of the resulting $2\NA\times 2\NA$ matrix is obtained as follows.

\medskip\noindent
\textbf{Block (even, even):}
\begin{align}
  \wv{\tilde{c}_{2m}\tilde{c}_{2n}}
  &= \wv{(\cdag_m + c_m)(\cdag_n + c_n)} \notag\\
  &= \wv{\cdag_m \cdag_n} + \wv{\cdag_m c_n} + \wv{c_m \cdag_n} + \wv{c_m c_n}
   = \mathcal{W}_4 + \mathcal{W}_3 + \mathcal{W}_2 + \mathcal{W}_1.
  \label{eq:block11}
\end{align}

\medskip\noindent
\textbf{Block (odd, odd):}
\begin{align}
  \wv{\tilde{c}_{2m-1}\tilde{c}_{2n-1}}
  &= \wv{i(\cdag_m-c_m)\cdot i(\cdag_n-c_n)} \notag\\
  &= -\wv{\cdag_m\cdag_n - \cdag_m c_n - c_m\cdag_n + c_m c_n}
   = -\mathcal{W}_4 + \mathcal{W}_3 + \mathcal{W}_2 - \mathcal{W}_1.
  \label{eq:block22}
\end{align}

\medskip\noindent
\textbf{Block (even, odd):}
\begin{align}
  \wv{\tilde{c}_{2m}\tilde{c}_{2n-1}}
  &= \wv{(\cdag_m + c_m)\cdot i(\cdag_n - c_n)} \notag\\
  &= i\wv{\cdag_m\cdag_n - \cdag_m c_n + c_m\cdag_n - c_m c_n}
   = i(\mathcal{W}_4 - \mathcal{W}_3 + \mathcal{W}_2 - \mathcal{W}_1).
  \label{eq:block12}
\end{align}

\medskip\noindent
\textbf{Block (odd, even):}
\begin{align}
  \wv{\tilde{c}_{2m-1}\tilde{c}_{2n}}
  &= \wv{i(\cdag_m - c_m)\cdot(\cdag_n + c_n)} \notag\\
  &= i\wv{\cdag_m\cdag_n + \cdag_m c_n - c_m\cdag_n - c_m c_n}
   = i(\mathcal{W}_4 + \mathcal{W}_3 - \mathcal{W}_2 - \mathcal{W}_1).
  \label{eq:block21}
\end{align}

\medskip\noindent
Putting all four blocks together (with even indices occupying the first $\NA$ rows/columns and odd indices the last $\NA$):
\begin{equation}
  \wv{\tilde{c}\tilde{c}}
  =
  \begin{pmatrix}
    \mathcal{W}_1+\mathcal{W}_2+\mathcal{W}_3+\mathcal{W}_4 & i(-\mathcal{W}_1+\mathcal{W}_2-\mathcal{W}_3+\mathcal{W}_4) \\[4pt]
    i(-\mathcal{W}_1-\mathcal{W}_2+\mathcal{W}_3+\mathcal{W}_4) & -\mathcal{W}_1+\mathcal{W}_2+\mathcal{W}_3-\mathcal{W}_4
  \end{pmatrix}.
  \label{eq:Gamma_matrix_full}
\end{equation}
We define the $2\NA\times 2\NA$ antisymmetric matrix $\Gamma$ via:
\begin{equation}
  \wv{\tilde{c}_m\tilde{c}_n} \;=\; \delta_{mn} + i\Gamma_{mn}.
  \label{eq:Gamma_def}
\end{equation}
Here $i\Gamma$ is the Majorana representation of $C^{1|2}$, restricted to subsystem $A$. Taking the weak value of the Majorana anti commutator $\{\tilde{c}_m,\tilde{c}_n\}=2\delta_{mn}$ gives $\wv{\tilde{c}_m\tilde{c}_n}+\wv{\tilde{c}_n\tilde{c}_m}=2\delta_{mn}$, which fixes the symmetric part of $\wv{\tilde{c}_m\tilde{c}_n}$ to $\delta_{mn}$.  All the non-trivial physical correlations, therefore reside in the antisymmetric part, which is isolated in $i\Gamma_{mn}$.  From equation \eqref{eq:Gamma_matrix_full} and equation \eqref{eq:Gamma_def}:
\begin{equation}
  i\Gamma
  \;=\;
  \wv{\tilde{c}\tilde{c}} - \mathbf{1}_{2\NA}
  \;=\;
  \begin{pmatrix}
    \mathcal{W}_1+\mathcal{W}_2+\mathcal{W}_3+\mathcal{W}_4-\mathbf{1} & i(-\mathcal{W}_1+\mathcal{W}_2-\mathcal{W}_3+\mathcal{W}_4) \\[4pt]
    i(-\mathcal{W}_1-\mathcal{W}_2+\mathcal{W}_3+\mathcal{W}_4) & -\mathcal{W}_1+\mathcal{W}_2+\mathcal{W}_3-\mathcal{W}_4-\mathbf{1}
  \end{pmatrix}.
  \label{eq:iGamma_explicit}
\end{equation} 

We now have two independent expressions for $\wv{\tilde{c}_m\tilde{c}_n}$. The first is obtained directly from the states $|\psi_1\rangle$ and $|\psi_2\rangle$ by computing the correlators $\mathcal{W}_1, \mathcal{W}_2, \mathcal{W}_3, \mathcal{W}_4$ and assembling the explicit matrix \eqref{eq:Gamma_matrix_full}, which can be diagonalised numerically to obtain the eigenvalues $\pm\nu_k$. The second follows from the diagonal ansatz \eqref{eq:tau_diagonal}; since $\wv{\tilde{c}_m\tilde{c}_n} = \mathrm{Tr}_A[\tilde{c}_m\tilde{c}_n\,\tauA]$, evaluating this trace in the diagonal mode basis gives the $2\times 2$ block structure \eqref{eq:iGamma_block_apx}, which tells us that the eigenvalues of $i\Gamma$ must have the form $\pm\tanh(\epsilon_k/2)$. Since both expressions describe the same matrix $i\Gamma$, matching their eigenvalues gives the relation $\nu_k = \tanh(\epsilon_k/2)$, which means the $\nu_k$ obtained by diagonalising the correlator matrix \eqref{eq:iGamma_explicit} can be substituted directly into the pseudo entropy formula, without ever needing to find $\epsilon_k$ explicitly.
The diagonal ansatz \eqref{eq:tau_diagonal} is written in the basis of decoupled modes $\{d_{A,k},\, d_{A,k}^\dagger\}$. We introduce the corresponding Majorana operators for these modes:
\begin{equation}
  \tilde{d}_{A,2k} = d_{A,k}^\dagger + d_{A,k},
  \qquad
  \tilde{d}_{A,2k-1} = i(d_{A,k}^\dagger - d_{A,k})
  \label{eq:d_majorana}
\end{equation}
In this basis, $\tauA = \bigotimes_k \tau_k$ where $\tau_k = e^{-\epsilon_k n_k}/(1+e^{-\epsilon_k})$ and $n_k = d_k^\dagger d_k$. Using relation \eqref{eq:d_majorana} we calculate diagonal and off diagonal elements as follows. We first compute $\mathrm{Tr}[\tau_k\,\tilde{d}^2_{A,2k}]$ for the diagonal block by expanding the Majorana operator squared:
\begin{align*}
  \tilde{d}_{A,2k}^2
  &= (d_{A,k}^\dagger + d_{A,k})(d_{A,k}^\dagger + d_{A,k}) \\
  &= (d_{A,k}^\dagger)^2 + d_{A,k}^\dagger d_{A,k} + d_{A,k}d_{A,k}^\dagger + (d_{A,k})^2 \\
  &= 0 + n_k + (1-n_k) + 0 = \mathbf{1}
\end{align*}
Hence, the expectation value with respect to $\tau_k$ becomes
\begin{align*}
  \mathrm{Tr}[\tau_k\,\tilde{d}_{A,2k}^2]
  = \mathrm{Tr}[\tau_k\,\mathbf{1}]
  = \mathrm{Tr}[\tau_k]
  = 1
\end{align*}

Therefore, diagonal blocks are 1: 
\begin{equation}
 \wv{\tilde{d}_{A,2k}\tilde{d}_{A,2k}} = 
  \wv{\tilde{d}_{A,2k-1}\tilde{d}_{A,2k-1}} = 1 
\end{equation}

Similarly, we now compute
$\mathrm{Tr}[\tau_k\,\tilde{d}_{A,2k}\tilde{d}_{A,2k-1}]$
for the off-diagonal block by expanding the product of Majorana operators:
\begin{align*}
  \tilde{d}_{A,2k}\,\tilde{d}_{A,2k-1}
  &= (d_{A,k}^\dagger + d_{A,k})\cdot i(d_{A,k}^\dagger - d_{A,k}) \\
  &= i\left[(d_{A,k}^\dagger)^2 - d_{A,k}^\dagger d_{A,k}
     + d_{A,k}d_{A,k}^\dagger - (d_{A,k})^2\right] \\
  &= i\left[0 - n_k + (1-n_k) - 0\right] \\
  &= i(1-2n_k).
\end{align*}

We now evaluate the trace with respect to $\tau_k = \frac{1}{1+e^{-\epsilon_k}}\begin{pmatrix}1&0\\0&e^{-\epsilon_k}\end{pmatrix}$ written in the occupation-number basis
$\{|0\rangle_k, |1\rangle_k\}$,
where $n_k|n\rangle = n|n\rangle$:
\begin{align}
  \mathrm{Tr}\!\left[\tau_k\cdot i(1-2n_{A,k})\right]
  &= \frac{i}{1+e^{-\epsilon_k}}
     \Bigl[1\cdot(1-2\cdot 0) + e^{-\epsilon_k}\cdot(1-2\cdot 1)\Bigr]
   = \frac{i(1-e^{-\epsilon_k})}{1+e^{-\epsilon_k}}.
  \label{eq:trace_off_diag}
\end{align}
Multiplying numerator and denominator by $e^{\epsilon_k/2}$:
\begin{align}
  \frac{i(1-e^{-\epsilon_k})}{1+e^{-\epsilon_k}}
  = \frac{i(e^{\epsilon_k/2}-e^{-\epsilon_k/2})}{e^{\epsilon_k/2}+e^{-\epsilon_k/2}}
  = i\tanh\!\left(\frac{\epsilon_k}{2}\right).
  \label{eq:tanh_simplification}
\end{align}
By antisymmetry of Majorana products
($\tilde{d}_{A,2k-1}\tilde{d}_{A,2k} = -\tilde{d}_{A,2k}\tilde{d}_{A,2k-1}$),
the other off-diagonal entry is:
\begin{equation}
  \wv{\tilde{d}_{A,2k}\tilde{d}_{A,2k-1}} = i\tanh\!\left(\tfrac{\epsilon_k}{2}\right),
  \qquad
  \wv{\tilde{d}_{A,2k-1}\tilde{d}_{A,2k}} = -i\tanh\!\left(\tfrac{\epsilon_k}{2}\right).
  \label{eq:off_diag_blocks}
\end{equation}
In the diagonal-mode basis, each $2\times 2$ block of
$\wv{\tilde{d}\tilde{d}}$ for mode $k$ is:
\begin{equation}
  \wv{\tilde{d}\tilde{d}}\big|_k
  = \begin{pmatrix} 1 & i\tanh(\epsilon_k/2) \\ -i\tanh(\epsilon_k/2) & 1 \end{pmatrix}.
  \label{eq:dd_block}
\end{equation}
Subtracting $\mathbf{1}$ according to equation \eqref{eq:Gamma_def}, each
$2\times 2$ block of $i\Gamma$ is:
\begin{equation}
  i\Gamma_k
  = \begin{pmatrix} 0 & i\tanh(\epsilon_k/2) \\ -i\tanh(\epsilon_k/2) & 0 \end{pmatrix}.
  \label{eq:iGamma_block_apx}
\end{equation}
The characteristic equation of the $2\times 2$ block \eqref{eq:iGamma_block_apx}
is:
\begin{equation}
  \det\begin{pmatrix}-\lambda & i\tanh(\epsilon_k/2) \\ -i\tanh(\epsilon_k/2) & -\lambda\end{pmatrix}
  = \lambda^2 - \tanh^2\!\left(\frac{\epsilon_k}{2}\right) = 0,
  \label{eq:char_eq}
\end{equation}
giving eigenvalues $\lambda = \pm\tanh(\epsilon_k/2)$.  Since the full $2\NA\times 2\NA$ matrix $i\Gamma$ is block-diagonal with blocks of the form \eqref{eq:iGamma_block_apx}, its $2\NA$ eigenvalues come in
$\NA$ pairs of opposite sign:
\begin{equation}
  \nu_k = \pm\tanh\!\left(\frac{\epsilon_k}{2}\right),
  \label{eq:nu_eps}
\end{equation}
We choose the $\NA$ positive eigenvalues $\nu_1,\ldots,\nu_{\NA}$.

\subsection{The Pseudo Entropy Formula}
\label{subsec:pseudo_entropy_formula}

From equation \eqref{eq:tau_diagonal}, each single-mode factor $\tau_k$ is a
$2\times 2$ matrix in the space $\{|0\rangle_k, |1\rangle_k\}$:
\begin{equation}
  \tau_k
  = \frac{e^{-\epsilon_k n_{A,k}}}{1+e^{-\epsilon_k}}
  = \begin{pmatrix}
      \dfrac{1}{1+e^{-\epsilon_k}} & 0 \\[8pt]
      0 & \dfrac{e^{-\epsilon_k}}{1+e^{-\epsilon_k}}
    \end{pmatrix}.
  \label{eq:tauk_matrix}
\end{equation}
We now rewrite the diagonal entries in terms of $\nu_k=\tanh(\epsilon_k/2)
=(e^{\epsilon_k}-1)/(e^{\epsilon_k}+1)$.  Adding 1 to $\nu_k$ and
dividing by 2:
\begin{equation}
  \frac{1+\nu_k}{2}
  = \frac{1}{2}\left(1+\frac{e^{\epsilon_k}-1}{e^{\epsilon_k}+1}\right)
  = \frac{1}{2}\cdot\frac{2e^{\epsilon_k}}{e^{\epsilon_k}+1}
  = \frac{e^{\epsilon_k}}{e^{\epsilon_k}+1}
  = \frac{1}{1+e^{-\epsilon_k}}.
  \label{eq:nu_upper}
\end{equation}
Subtracting $\nu_k$ from 1 and dividing by 2:
\begin{equation}
  \frac{1-\nu_k}{2}
  = \frac{1}{2}\left(1-\frac{e^{\epsilon_k}-1}{e^{\epsilon_k}+1}\right)
  = \frac{1}{2}\cdot\frac{2}{e^{\epsilon_k}+1}
  = \frac{1}{e^{\epsilon_k}+1}
  = \frac{e^{-\epsilon_k}}{1+e^{-\epsilon_k}}.
  \label{eq:nu_lower}
\end{equation}
Therefore:
\begin{equation}
  \tau_k
  = \begin{pmatrix}
      \dfrac{1+\nu_k}{2} & 0 \\[8pt]
      0 & \dfrac{1-\nu_k}{2}
    \end{pmatrix},
  \label{eq:tau_k_nu}
\end{equation}
with eigenvalues $\lambda_{k,0}=(1+\nu_k)/2$ (mode empty) and
$\lambda_{k,1}=(1-\nu_k)/2$ (mode filled).

The pseudo entropy is:
\begin{align}
  S\!\left(\tauA\right)
  &= -\mathrm{Tr}_A\!\left[\tauA\log\tauA\right] \notag\\
  &= -\sum_{k=1}^{\NA}\mathrm{Tr}\!\left[\tau_k\log\tau_k\right] \notag\\
  &= -\sum_{k=1}^{\NA}
     \bigl(\lambda_{k,0}\log\lambda_{k,0} + \lambda_{k,1}\log\lambda_{k,1}\bigr),
  \label{eq:PE_sum_modes}
\end{align}
where the second line uses $\tauA=\bigotimes_k\tau_k$ and the identity
$\log(\tauA)=\sum_k\mathbf{1}\otimes\cdots\otimes\log\tau_k\otimes\cdots\otimes\mathbf{1}$
together with $\mathrm{Tr}[\tau_l]=1$ for all $l$.
Substituting equation \eqref{eq:tau_k_nu}, we obtain the final result:
\begin{equation}
    S\!\left(\tau^A\right)
  \;=\;
  -\sum_{k=1}^{N_A}
  \left[
    \frac{1+\nu_k}{2}\,\log\frac{1+\nu_k}{2}
    +
    \frac{1-\nu_k}{2}\,\log\frac{1-\nu_k}{2}
  \right],
  \label{eq:PE_final}
\end{equation}
where $\pm\nu_k$ are the $2N_A$ eigenvalues of the matrix $i\Gamma$ defined in equation \eqref{eq:Gamma_def}, and we take the $N_A$ eigenvalues with positive  real part. The $\nu_k$ are generally complex when $|\psi_1\rangle\neq|\psi_2\rangle$, so $S(\tau^A)\in\mathbb{C}$ in general. Since the arguments $\tfrac{1\pm\nu_k}{2}$ are generically complex, the 
logarithms are evaluated on the principal branch $-\pi \;<\; \mathrm{Im}[\log z] \;\leq\; \pi,$
which is well-defined away from the branch cut on the negative real axis. The R\'enyi pseudo entropy is:
\begin{equation}
  S^{(n)}\!\left(\tauA\right)
  \;=\;
  \frac{1}{n-1}\sum_{k=1}^{\NA}
  \log\!\left[
    \left(\frac{1+\nu_k}{2}\right)^{\!n}
    -
    \left(\frac{1-\nu_k}{2}\right)^{\!n}
  \right].
  \label{eq:Renyi_PE}
\end{equation}
In the limit $n\to 1$, above equation recovers equation \eqref{eq:PE_final}. For $n=2$, the R\'enyi pseudo entropy becomes
\begin{align}
S^{(2)}(\tauA)
&=\frac{1}{2-1}
\sum_{k=1}^{N_A}
\log\!\left[
\left(\frac{1+\nu_k}{2}\right)^2
-
\left(\frac{1-\nu_k}{2}\right)^2
\right] \nonumber \\
&=
\sum_{k=1}^{N_A}
\log\!\left(
\frac{4\nu_k}{4}
\right)
\nonumber \\
&=
\sum_{k=1}^{N_A}\log(\nu_k).
\end{align}

Equivalently, $S^{(2)}(\tau_A)
=
\log\!\left(
\prod_{k=1}^{N_A}\nu_k
\right)$ which is the log of product of $\nu_k$'s. This can be useful for numerical calculations when pseudo-entropy becomes unstable. When $|\psi_1\rangle=|\psi_2\rangle$, the weak values reduce to ordinary expectation values, the $\nu_k$ become real with $\zeta_k\in[-1,1]$, and setting $\zeta_k=(1+\zeta_k)/2\in[0,1]$ the formula \eqref{eq:PE_final} reduces to the standard entanglement entropy $S(\rho_A)=-\sum_k[\zeta_k\log\zeta_k+(1-\zeta_k)\log(1-\zeta_k)]$.

\end{document}